\documentclass[aps,prd,twocolumn,showpacs,groupedaddress,preprintnumbers,amsmath,amssymb]{revtex4-1}  
\usepackage{graphicx}  
\usepackage{dcolumn}   
\usepackage{bm}        
\usepackage{amssymb}   
\usepackage{mathrsfs}
\usepackage{amssymb,amsmath}
\usepackage{latexsym}
\usepackage{cancel}
\usepackage[utf8]{inputenc}
\usepackage{hyperref}
\usepackage[titletoc, toc]{appendix}

\begin{document}



\title{Time-domain analysis of a dynamically tuned signal recycled interferometer for the detection of chirp gravitational waves from coalescing compact binaries}
\author{D. A. Simakov}
\affiliation{Max Planck institute for Gravitational Physics, Leibniz University of Hanover, Hanover,
30167, Germany} \email{dmitry.simakov@aei.mpg.de}
\date{\today}

\date{\today}

\begin{abstract}
In this article we study a particular method of detection of chirp signals from coalescing compact binary stars---the so-called dynamical tuning, i.e.,~amplification of the signal via tracking of its instantaneous frequency by the tuning of a signal-recycled detector. The motion of the signal-recycling mirror, the position of which defines the tuning of the detector, causes nonstationarity of the detector. The dynamically tuned detector can be simulated in a quasistationary approximation if the mirror position, amplitude, and frequency of a chirp signal are changing slowly.  A time-domain consideration developed for signal-recycled interferometers, in particular GEO\,600, describes the signal and noise evolution in the more general case of a purely nonstationary detector. We prove that the shot noise from the dark port and optical losses remains white in this case. The analysis of the transient effects shows that during the perfect tracking of the chirp frequency only transients from fast amplitude changes arise because the transients from changes of the detector tuning and signal frequency completely cancel each other. The slow change of the amplitude in this case establishes a so-called {\it virtually stationary} detection, meaning the signal fields at the detector hold their stationary values at each instance of time, corresponding to the instantaneous parameters of the gravitational wave and of the detector. The signal-to-noise-ratio gain from the implementation of dynamical tuning, calculated in this paper, is $\sim 17$ for a shot noise- limited GEO\,600-like detector and $\sim 7$ for a detector with both shot and displacement noise.
\end{abstract}

\maketitle 

\section{Introduction}

In the last few decades a big effort has been made to detect gravitational waves (GWs) from various sources in deep space. In particular, we expect a very interesting kind of GW signal, usually referred to as a chirp signal [see Fig.~\ref{fig:gw}(a) and \ref{fig:gw}(b) for an example], to be emitted by compact binary systems, such as a pair of neutron stars or black holes inspiraling toward each other and then  coalescing.
 
A chirp signal gives us unique information about nonlinear dynamics of matter and space-time, as the GWs are emitted from the regions with strong space-time curvature. Compact binary coalescence (CBC) and the corresponding GW signal are conventionally split into three stages: inspiral, merger, and ringdown. The post-Newtonian approximation of general relativity (GR) \cite{santamaria10, ajith11, kamaretsos12, kamaretsos12b} allows a precise prediction of most of the inspiral stage. At this stage, the signal has a sinusoidal shape with frequency and amplitude increasing in time. The latter stages of the inspiral and all of the merger and the ringdown stages are modeled by numerical relativity, and then all stages are continuously sewn together. 

Once a signal is measured and compared to the templates, one can extract 
information about masses and spins of the inspiraling binary objects as well 
as the equation of state of dense nuclear matter in the case of the merging 
neutron stars \cite{damour12, pannarale11}. Therefore, a sensitive detection 
of chirp signals might verify or falsify GR or alternative theories of 
gravity via comparing their predictions with the measured parameters. 
 Schutz in Refs.~\cite{schutz86} and later Taylor {\it et al.} in Ref.~\cite{taylor12} also proposed that the Hubble constant can be independently determined in a new and potentially 
 accurate way by observation of the inspiral stage of the chirp GWs.

\begin{figure}
  \includegraphics [angle=0,width=\columnwidth]{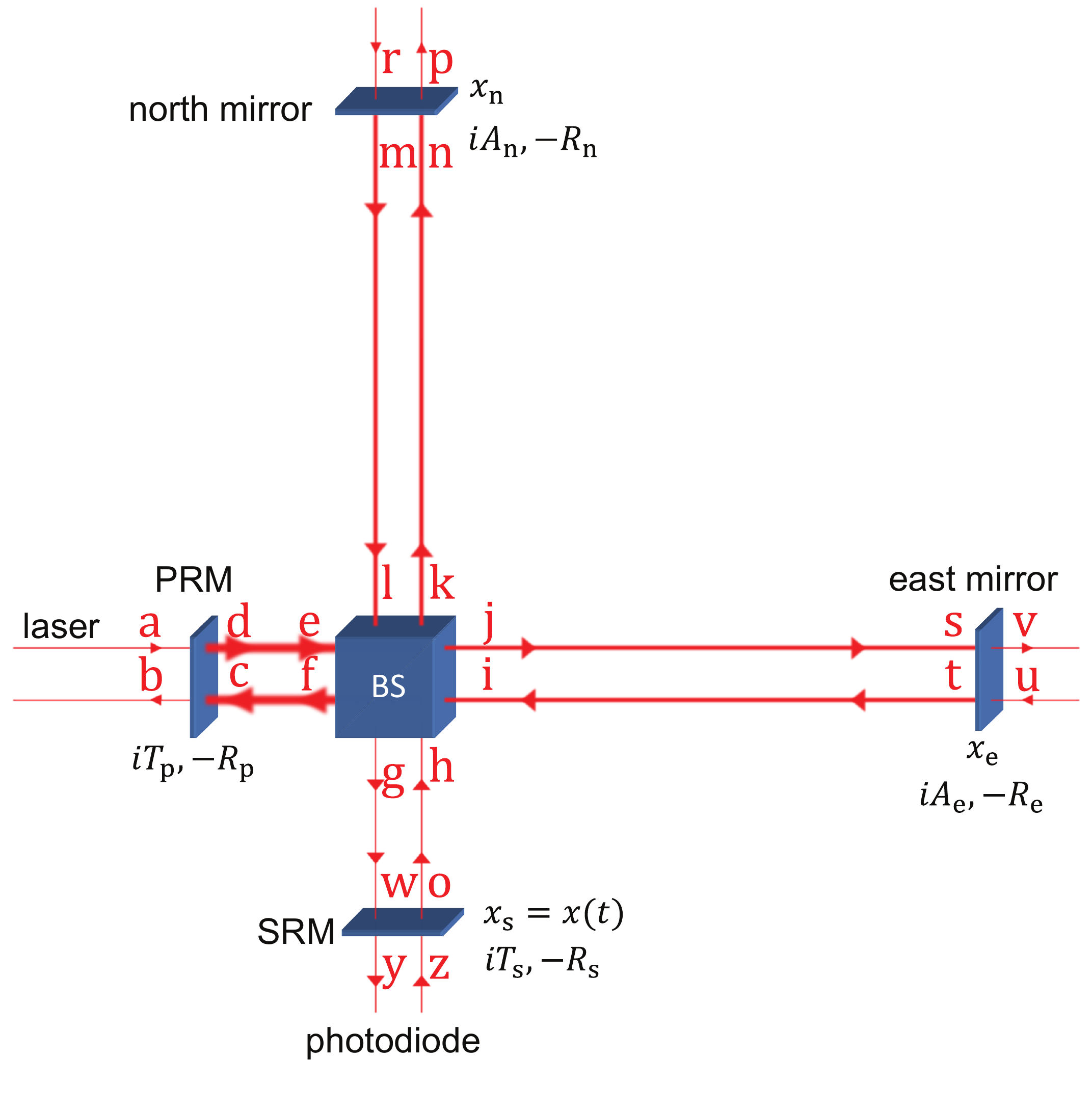}
  \caption{Scheme of the considered GW detector. The notations are presented in Table \ref{tab:eq}.}\label{fig:geo1}
\end{figure}

Nowadays, large-scale ground-based laser interferometers are the most sensitive detectors of GWs in the frequency range 10 Hz--5 kHz \cite{blair121}. Currently, the first generation of GW detectors has finished their operation without any detection, which agrees with the current estimations for their detection rate. The significantly improved sensitivity of the second-generation detectors will allow us to achieve a detection rate of about 25--400 $\text{yr}^{-1}$ \cite{shaughnessy10}. Upon reaching the Earth, the GWs are only tiny perturbations of the space-time metric causing small variation of the proper distances between the quasi-free-falling test masses of the laser interferometer. All currently operating and future planned GW detectors are based on the traditional Michelson topology, a typical idealized example of which is considered in this paper (see Fig.~\ref{fig:geo1}): the interferometer consists of a 50/50 beam splitter, perfectly reflecting end mirrors and additional mirrors for signal and power amplification, referred to as the signal-recycling mirror (SRM) and power-recycling mirror (PRM), respectively. Interferometers usually operate near the dark fringe in the output port, meaning that the laser beams reflected from the end mirrors destructively interfere on the beam splitter toward the photo diode. A GW of appropriate polarization and direction causes antisymmetric (differential) motion of the interferometer's end mirrors relative to the beam splitter. This breaks the destructive interference at the output port allowing a tiny part of the optical field carrying the information about the GW signal to reach the photodetector. This signal field gets recirculated by the SRM, forming the differential mode of the interferometer in the effective signal-recycling cavity (SRC). The PRM in the laser port increases power at the end mirrors and by recirculating the light reflected from them it creates the common mode of the interferometer in the power-recycling cavity (PRC). This mode is sensitive to the symmetric (common) motion of the end mirrors. Therefore the common mode of the detector does not contain any information about the GW signal and in the rest of this paper we only consider the differential mode.

\begin{figure}
  \includegraphics [angle=0,width=\columnwidth]{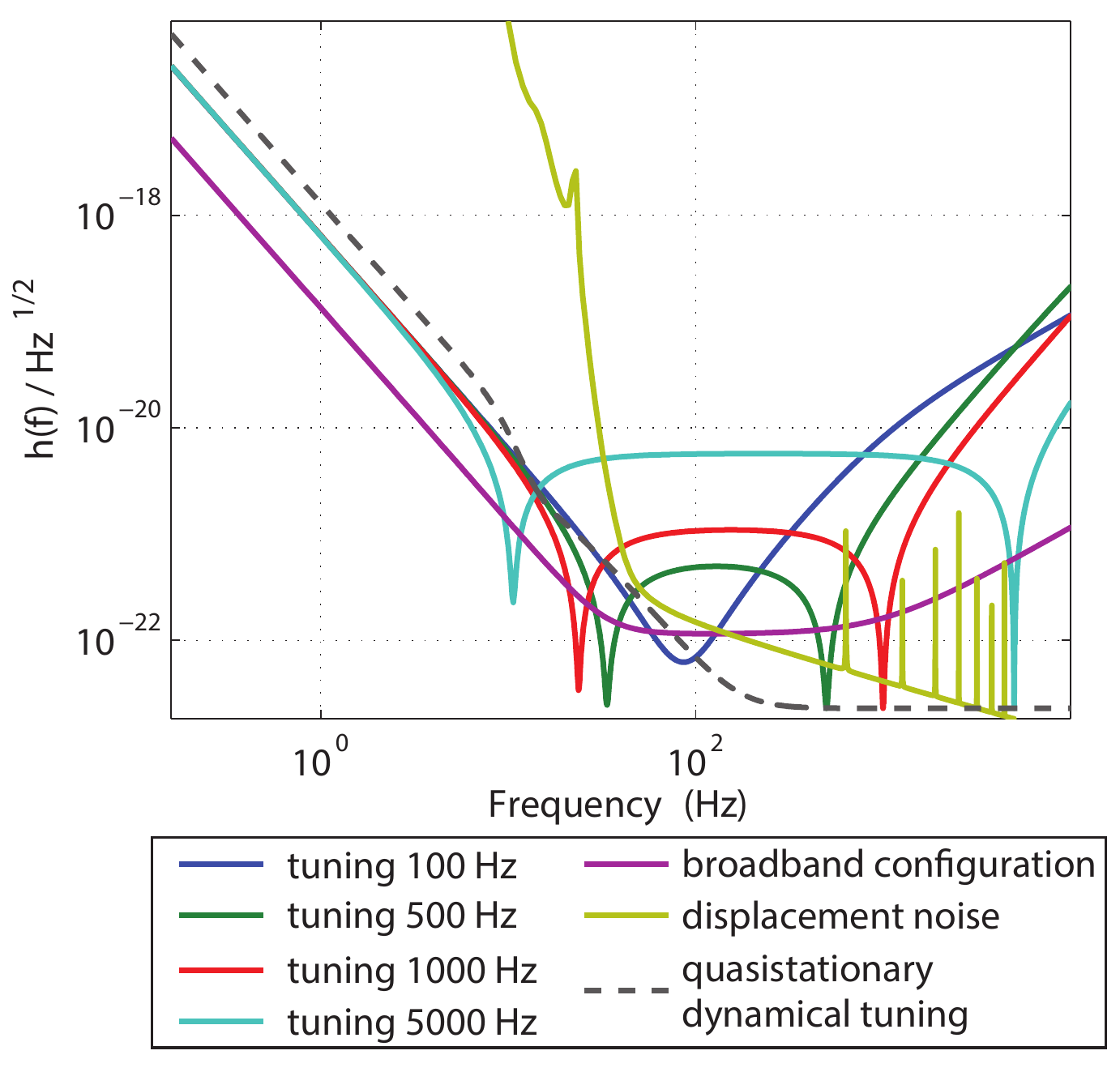}
  \caption{The quantum noise of broadband and narrow band detector configurations. The quantum noise of a quasistationary dynamical tuning (the points of optical resonance in curves corresponding to each tuning) is also presented to compare with the mirror displacement noise.}\label{fig:bn}
\end{figure}

Parameters of the SRC are determined by the properties of the SRM: the frequency bandwidth of the cavity is determined by the SRM transmittance, and the detuning of the laser carrier frequency from cavity resonance is determined by the microscopic position of the SRM. In  this sense, the SRC is equivalent to a simple Fabry--Perot cavity \cite{buonnano02}. The SRC can be tuned to any desired signal frequency via the proper choice of the cavity detuning. Currently, all GW detectors operate stationary in time, meaning that the parameters of the SRC are fixed. There are two typical regimes of detection of chirp signals in this case: a broadband operation and a narrow band operation (see Fig.~\ref{fig:bn}). In the former regime the detector is sensitive to the entire frequency band of the chirp signal, but at moderate sensitivity. On the contrary, in the narrow band regime, the detector is much more sensitive, but only in a narrow band around the signal frequency to which the SRC is tuned (see Fig.~\ref{fig:bn}). Since the chirp signal at the inspiral stage is a sine function with frequency increasing in time [see Fig.~\ref{fig:gw}(b)], the peak sensitivity of the narrow-band-operated detector will only be achieved during the short interval of time, when the particular instantaneous frequency of the chirp approximately coincides with the detuning of the SRC.

Another option for the detection of a chirp signal was proposed by Meers {\it et al.} in Ref. \cite{meers93}: real-time tuning of a narrow-band SRC to the instantaneous frequency of the signal via positioning of the SRM, i.e.,~real-time signal tracking. This method of detection is referred to as {\it dynamical tuning}. However, analysis in Ref. \cite{meers93} was performed under the following approximations: (i) a shot-noise-limited detector, and (ii) slow enough motion of the SRM such that the detector can be considered  as a quasistationary one; i.e.,~all the optical fields evolve adiabatically on the time scale of the motion of the SRM. The latter approximation also sets the limiting instant of time until which the signal can be observed before entering the regime of rapid frequency increase, where quasistationary approximation does not hold anymore. To agree with these approximations, the authors considered detecting only a part of the chirp signal---with the instantaneous frequency varying from 100 up to 500\,Hz. The method developed in this paper allows us to treat the problem of dynamical tuning outside of these approximations. It should be noted that we do not consider the problem of signal prediction; we assume that the initial time evolution of the signal is known, for instance, from the low-frequency data of other GW detectors, such that the subsequent evolution of the signal can be predicted.

\begin{figure}
  \includegraphics [angle=0,width=\columnwidth]{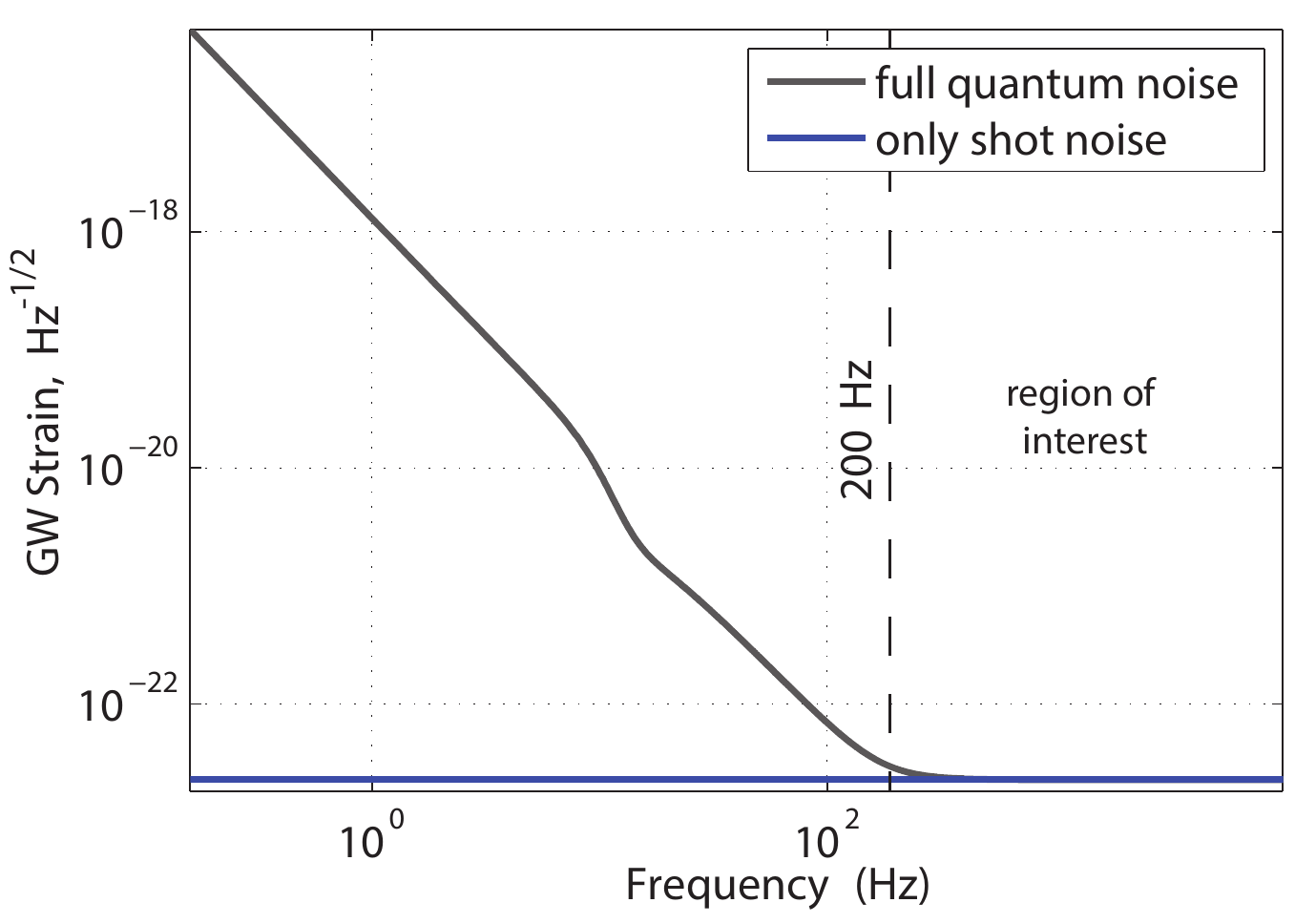}
  \caption{The quasistationary approximations of dynamical tuning for the detector with the full quantum noise (with radiation pressure) and with the shot noise only.}\label{fig:quasi}
\end{figure}

The response of a stationary-operated detector to GWs and all kinds of noise sources is usually calculated in the frequency domain. For a detailed analysis, see Refs. \cite{blair121, KLMTV, levin97, levin08, harry02, Liu00}. A GW detector performing dynamical tuning operates in the nonstationary regime. The tracking of a chirp signal with a slowly changing frequency may be described with a quasistationary approximation, assuming the detector reaches steady state very fast. A quasistationary approximation can also be considered in the frequency domain, as it was performed in Ref. \cite{meers93}. However, when the frequency of the signal and, correspondingly, position of the SRM change too fast, a frequency-domain analysis is not adequate, and therefore we develop a time-domain analysis to model it properly. In particular, the detector response takes the form of a series over an infinite number of round trips of light inside the SRC \cite{offrein94, lawrence99}, the so-called impulse response. 

The basics of time-domain and frequency-domain analyses of laser GW detectors this paper is grounded on, including the stationary responses to common and differential modes of a stationary operating interferometer, and its shot-noise sensitivity formulas, are given with sufficient details in Ref. \cite{rakhmanov00}.  

Using our time-domain model, we also calculate the response of the detector to shot noise (vacuum fluctuation of the electromagnetic field injected from the dark port or lossy optical elements) and to differential motion of the end mirrors (caused by the GWs and various types of mirror-displacement noise such as thermal noise). The analysis of the transient processes during ideal dynamical tuning, performed using this approach, has shown that only fast amplitude changes cause a deviation from the quasistationary predictions, while the transients from signal frequency and from the SRM position cancel each other. The slow change of amplitude establishes a so-called {it virtually stationary} detection, when the output signal at every instance of a nonstationary detection has the stationary value, corresponding to the instantaneous parameters of the signal and of the detector.  The radiation pressure noise (back action) is omitted for the considered power at the end mirrors in the current model because (i) its typical frequencies are lower than the characteristic frequencies of the considered part of chirp signals starting from 200\,Hz (see Fig.~\ref{fig:quasi}) and  (ii) it is dominated by the other noise sources (see Fig.~\ref{fig:bn}). 

Finally, we study the possible signal-to-noise-ratio (SNR) gains from the implementation of dynamical tuning to the traditional broadband stationary operated detector. For the shot-noise-limited detector the increase of the SNR is $\sim 17$, and for the detector with both displacement and shot noise, the increase is $\sim 7$. We found that, in contrast to the stationary operated detector limited by both displacement and shot noise, the detector performing dynamical tuning is only displacement noise limited. This happens because displacement noise and GW signal, both creating the differential motion of the end mirrors, are resonantly enhanced by dynamical tuning in the same manner (effectively, dynamical tuning tracks and amplifies the same components of displacement noise as of the GW signal), while shot noise on the photodetector remains the same (more precisely, shot noise remains delta correlated independently of the motion of the SRM).

The paper is organized as follows.  We derive the time-domain response of the detector to a gravitational wave in Sec. \ref{detRespGW}. We consider the influence of displacement noise on the detector output in Sec. \ref{detRespDN} and by the shot noise in Sec. \ref{detRespSN}. The time-domain models are presented in these sections with a consistency check in comparison to the stationary model. The features of a nonstationary model and, especially, its difference from a quasistationary approximation are represented in Sec. \ref{DynBeh}. In Sec. \ref{simRes} we present the SNR gain with respect to a stationary detector, which can be achieved with dynamical tuning.

\section{Model for the signal induced by a GW during dynamical tuning}\label{detRespGW}

\begin{table*}
	\caption{Notations and definitions used in this paper.}
	\centering
	\begin{tabular}{c l}
		\hline
		Notation        &definition                 \\
		&                             \\
		\hline
		\hline
		
		$\omega_{\rm p}$  &Frequency of the carrier laser     \\
		$c$			&Speed of light                          \\
		$\mathcal{A} $          & Cross-section of the detected beam                           \\               
		$ R_{\rm f}  $           & Equivalent end-mirror reflectivity \eqref{eqRef}                           \\               
		$T_{\rm s}	$			& Transmittance of the SRM \\
		$R_{\rm s}	$			& Reflectivity of the SRM  \\
		$E$						& Field falling on the beam splitter (see Fig.~\ref{fig:geo1})\\
		$\phi_{\rm e}$			& Phase of the field $E$\\
		$\phi_{\rm lo}$			& Local oscillator of the homodyne detector\\
		$\phi_{\rm h}$				& Homodyne angle \eqref{hangle}\\
		$k_{\rm p}$				& Wave vector $\frac{\omega_{\rm p}}{c}$ \\
		$x(t)$					& Microscopic displacement of the SRM from the resonant position\\
		$\tau$					& Round-trip time \eqref{rt} \\
		\hline 
	\end{tabular}
	\label{tab:eq}
\end{table*}

Let us consider the GW detector described above. A plus-polarized gravitational wave $h(t)$, falling perpendicular onto the detector, causes  a differential motion of the end mirrors proportional to the arm length, assuming they have initially rested,
\begin{equation}\label{GW2length}
x_{\rm d}(t) = \frac{L h(t)}{2},
\end{equation}
where 
\begin{equation}\label{difx}
x_{\rm d}(t) = \frac{x_{\rm e}(t)-x_{\rm n}(t)}{2}
\end{equation}
is the differential motion, and $x_{\rm e}(t)$ and $x_{\rm n}(t)$ are the displacements of the east and the north end mirrors, respectively (see Fig.~\ref{fig:geo1}). $L$ is the arm length.

The differential displacement of the end mirrors breaks the dark-port condition of a Michelson interferometer, injecting signal sidebands into the SRC. In this paper, we neglect the optomechanical back-action effects, so the further evolution of sidebands may be reduced to the two elementary phenomena: (i) reflections (and transmissions) from the mirrors and a beam splitter and (ii) propagation through space. The optical losses in our case may be effectively reduced to the reflections. As part of this evolution, some field exits the dark port and is detected by a photodetector. The resulting output current on the photodetector $I_{\rm y}(t)$ contains the information about the differential motion of the end mirrors, caused by gravitational waves. 

The response of the detector to the signal under these circumstances can be considered linear for two reasons:  (i) the smallness of the end-mirror displacements caused by the GW in comparison to the wavelength of the light and (ii) the superposition principle for the fields inside the cavity. The output current of such a linear system in the time domain is defined by the end-mirror input motion via a so-called impulse response $L_{\rm s\rightarrow c}(t,t_1)$,
\begin{equation}\label{linres}
I_{\rm y}(t) = \int\limits_{-\infty}^{t}L_{\rm s\rightarrow c}(t, t_{ 1})x_{\rm d}(t_{1})dt_1,
\end{equation}
where $\rm s\rightarrow c$ stands for ``signal to current." Thus, once we know the linear response of the detector, we can simulate the output current for any gravitational wave.

\begin{figure}
  \includegraphics [angle=0,width=\columnwidth]{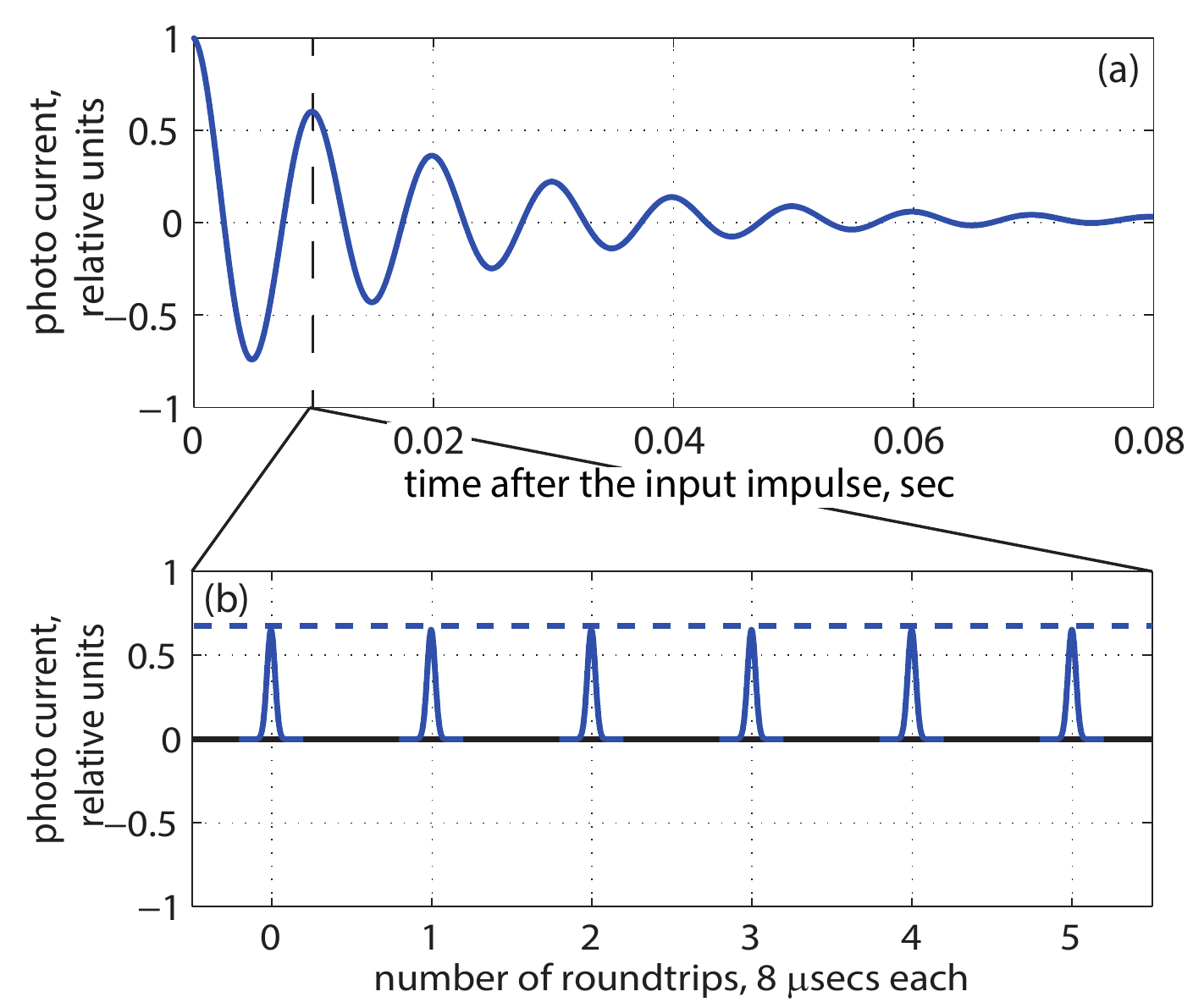}
  \caption{The typical impulse response of the considered detector with a constant detuning $f_{\rm tun} = 100$\,Hz: (a) in the response decay-time scale and (b) in the single round-trip time scale ($\sim 8 \mu$s).}\label{fig:imp}
\end{figure}

The physical meaning of the impulse response of the detector is the photocurrent caused by delta-impulse-shaped differential jitter of the end mirrors in the instance of time $t_1$. It contains the information about the signal transformation inside the detector; hence, all the optical parameters as well as the motion of the SRM are encrypted in it. 

The impulse response for GEO\,600 with a moving SRM has the following explicit expression:
\begin{multline}\label{impulses2pd}
L_{\rm s\rightarrow c}(t,t_1)  = \sum\limits_{n=0}^{\infty}C_{n}\cos[\xi_n(t)]\times \\
\times \delta\left(t_1-t+n\tau + \frac{\tau}{2}\right).
\end{multline}
Physically, it is a sequence of the output pulses with amplitudes $C_n$ and phases $\xi_n(t)$ describing a decaying and oscillating envelope. The pulses occur every round trip $\tau$ in the output photocurrent. The expressions for amplitudes and phases are explicitly presented in Eq. \eqref{2pddet}. 

The notations and definitions of the physical values used in the expression for an impulse response (and in other models of this paper) are presented in Table \ref{tab:eq}. Its derivation is shown in Appendix \ref{ssec:GEO_dif}, while the description of the time-domain model this derivation is based on can be found in Appendix \ref{app:GEO}. The basic features of the model are considered in the simpler case of a Fabry--Perot cavity in Appendix \ref{app:FP}.

The depiction of the impulse response is presented in Fig.~\ref{fig:imp} for a 100 Hz detuning in two scales: of seconds and of single pulses ($\sim 8 \mu$s). The typical set of parameters of GEO\,600, used for the simulations of this plot (as well as for other calculations in this paper), is presented in Tables \ref{tab:GEOconst} and \ref{tab:GEOvar}. The tables also include the current parameters of GEO\,600 and describe the changes required for the implementation of dynamical tuning.   

The physics standing behind the impulse response is the following. The signal pulse created in the SRC by the differential end-mirror movement makes the round trips between the SRM and the end mirrors with the period
\begin{equation}\label{rt}
	\tau = 2 \frac{L}{c}.
\end{equation}
Every time the pulse is reflected from the SRM the small part of it leaks from the cavity and is detected at the homodyne detector with the local oscillator (LO) field \eqref{LO}. 

These leaked pulses form the infinite number of decaying ``echoes" at the output with amplitudes $C_0, C_1, C_2, ... $. Two consequent pulses in this sequence are differing by a decay factor during one round trip $R_{\rm s} R_{\rm f}$, and the phase shift between two consequent impulses is obtained during the reflection from the SRM due to its microscopic displacement from the resonant position.

The reflection of light from the beam splitter and from the end mirrors is equivalent to the reflection from a single mirror with the reflectivity $R_{\rm f}$ defined in Eq. \eqref{eqRef}. This mirror and the SRM form an equivalent Fabry--Perot cavity. A more thorough description of the equivalence between the SRC of GEO\,600 topology and a single Fabry--Perot cavity is presented in Appendix \ref{equiv}.


The impulse response \eqref{impulses2pd} describes the behavior of a detector also in the stationary case when $\xi_n(t) = {\rm const}$. The Fourier transformation turns the impulse response into transfer function $R_{\rm s \rightarrow c}(\Omega)$ presented in Eq. \eqref{TransFuncX}. The transfer function describes the response of the detector on the sine gravitational wave. The wave creates two sidebands in the detector, the amplification of which has resonant features. The transfer function $R_{\rm s \rightarrow c}(\Omega)$ coincides with those, obtained conventionally in the frequency domain, e.g.,~in Refs \cite{rakhmanov00, Harms03, danilishin12, simakov14}. Therefore, the model for the time-domain response on the gravitational wave is consistent with the accepted frequency-domain models.

\section{Model for the end-mirror displacement induced noise during dynamical tuning}\label{detRespDN}
Gravitational waves cause differential end-mirror motion. However, it is not the only source of this motion. There is a number of stochastic influences on the end mirrors in GEO\,600 causing it, the most significant of which are thermal noise in mirror coating, seismic noise, and gravity gradient noise \cite{geonoise, har}. The noise is stationary and characterized by its spectral density $S(\Omega)$, the theoretical prediction of which is known and is depicted in Fig.~\ref{fig:noises}.

The impulse response to the differential end-mirror motion \eqref{impulses2pd} defines the output current at the photodetector for an arbitrary input. It is also applicable to stochastic motion of the end mirrors. 

For stationary detectors, it does not matter whether noise and signals are compared at the end mirrors or at the output photocurrent. The reason is that a linear detector transforms the same frequency components of both noise and signal equally, so the ratio of the intensities of these components does not change. Something similar happens for the nonstationary detector. Although the frequency components here do not evolve independently, the impulse response transforms the same components of the signal as of the noise in the same way. So, if the information about the signal is transferred completely, it also does not matter where one calculates the sensitivity with respect to the displacement noise. The completeness of the transferred signal can be proven by consequent action of the direct and of the inverse  impulse response, as it is shown in Eq. \eqref{invMatDet}.

When we investigate the influence of displacement noise on the sensitivity of the dynamically tuned detection, we consider displacement noise alone, neglecting shot noise. In this case, it is more convenient to compare the signal and noise at the end mirrors for the following reasons. The first is the convenient shape of the signal: it is proportional to the gravitational wave strain (assuming the initial position and velocity of the end mirrors equal zero). The second reason is stationarity of the displacement noise, simplifying the calculations. The third reason is the independence of the ratio between the signal and the displacement noise from the motion of the SRM required for the dynamical tuning.  So the SNR may be calculated in the conventional way, e.g.,~in the frequency domain, using the expression
\begin{equation}\label{SNR_disp}
d^2=\int\limits_{-\infty}^{\infty} \frac{\left|\tilde{x}_{\rm d}(\Omega)\right|^2}{S(\Omega)}d\Omega,
\end{equation}
where $\tilde{x}_{\rm d}(\Omega)$ is the Fourier transform of the signal differential mirror motion $x_{\rm d}(t)$.

If we want to calculate a more realistic sensitivity including both displacement and shot noise, we should treat their influences at the same part of the interferometer. For this purpose it may be more convenient to treat displacement noise at the output of the detector, i.e.~in the signal photocurrent. As it was mentioned, the autocorrelation function of this noise may be found from the one of the end-mirror motion noise via the impulse response \eqref{impulses2pd},
\begin{multline}\label{outth}
B_{\rm th}(t_1,t_2) = \sum\limits_{m=0}^{\infty}\sum\limits_{n=0}^{\infty}C_{m}C_n\cos\xi_m(t_1) \cos\xi_n(t_2) \times \\
\times B(t_2-t_1+(m-n)\tau),
\end{multline}
where $B(\tau)$ can be found from the spectral density $S(\Omega)$ mentioned above.

\section{Model for the shot-noise evolution during dynamical tuning}\label{detRespSN}

\subsection{Fields in the detector}

Quantum shot noise is conventionally considered as ground-state quantum oscillations injected into a cavity through any open port and from lossy elements \cite{KLMTV}. Since Maxwell's equations are valid for quantum mechanics, the quantum operator of the electromagnetic field can be treated like the classical field values in the previous sections. 

Let us consider, for example, the quantum field at the point ${\rm a}$, where the laser shines into the detector (see Fig.~\ref{fig:geo1}). The quantum electrical field operator in this point, describing also the classical part of the light when it is required, reads
\begin{multline}\label{em}
\hat{E}_{\rm a}(t)=\int\limits_0^{\infty}\sqrt{\frac{2 \pi \hbar\omega}{\mathcal{A}c}}[\hat{\rm a}(\omega)e^{-i\omega t}+\hat{\rm a}^+(\omega)e^{i\omega t}]\frac{d\omega}{2\pi} \approx \\
\approx  \sqrt{\frac{2 \pi \hbar\omega_{\rm p}}{\mathcal{A}c}}[\hat{\rm a}(t)e^{-i\omega_p t}+\hat{\rm a}^+(t)e^{i\omega_{\rm p} t}], 
\end{multline}
where $\hat{\rm a}(\omega)$ and $\hat{\rm a}^+(\omega)$ are the annihilation and creation operator, and $\mathcal{A}$ is the effective optical cross section of the considered beam. The annihilation and creation operators in the other spatial points of GEO\,600 are denoted by the corresponding letter depicted in Fig.~\ref{fig:geo1}. The operator $\hat{\rm a}^(t)$, introduced in \eqref{em}, is a Fourier transform of the annihilation operator $\hat{\rm a}(\omega)$, and represents the amplitude of the electric field. One has to note, that this Fourier transform is performed only within the frequency of the anticipated GWs, which is much smaller than the frequency of the laser carrier:
\begin{equation}\label{omapp}
\Omega \equiv \omega-\omega_{\rm p} \ll \omega_{\rm p}.
\end{equation}

In this section, from all the points in Fig.~\ref{fig:geo1}  we are only interested in the shot-noise injections from the dark-port $\hat{\rm z}(t)$, injections from the end mirrors $\hat{\rm r}(t)$ and $\hat{\rm u}(t)$, and the detector dark-port output $\hat{\rm y}(t)$.

The homodyne detection of the output field $\hat{E}_{\rm y}(t)$ gives a signal in the photocurrent, in which we can read out the GWs and see the noise. The main part of shot noise at the output is formed from the ground-state oscillations injected from the dark-port $\hat{\rm z}(t)$. The homodyne detection of this input field with the local oscillator \eqref{LO} results in a white spectrum for its noise in the frequency band of detection,
\begin{equation}\label{SNInp}
B_{\rm in}(t_1,t_1) = C_{\rm z}\delta(t_1-t_2),
\end{equation}
where $C_{\rm z}$ is a constant, determined by the amplitude of the local oscillator, and is explicitly presented in \eqref{SNConst}.

\subsection{Output field expressed in terms of input}
The output shot noise is the result of the evolution of the injected ground-state oscillations $\hat{E}_{\rm z}(t)$. This evolution, consisting of a phase shift, propagation, and amplitude change, is described by the complex amplitudes. Therefore, we can express the amplitude of the output field in terms of the input using the complex impulse response,
\begin{equation}\label{outint}
\hat{\rm y}(t) =\int\limits_{-\infty}^{t}L_{\rm c}(t, t')\hat{\rm z}(t')dt',
\end{equation}
where $L_{\rm c}(t,t_1)$ is the complex impulse response and $\hat{\rm y}(t)$ and $\hat{\rm z}(t)$ are the amplitudes of the output and input fields, respectively (see Fig.~\ref{fig:geo1}). The physical meaning of this function is a response on the infinitely short pulse with the optical carrier frequency. 

The noise we get by the homodyne detection of this field with the local oscillator \eqref{LO}, keeping in mind \eqref{SNInp}, reads
\begin{multline}\label{vacautocor}
B_{\rm vac}(t_1,t_2) = C_{\rm z}\int\limits_{-\infty}^{\min(t_1,t_2)}dt_1'\times \\
\times  \Re\left(L_{\rm s}(t_1,t_1')L_{\rm s}^*(t_2,t_1')\right).
\end{multline}
Here,
\begin{equation}\label{auximp}
L_{\rm s}(t,t_1) = L_{\rm c}(t, t_1)e^{i\omega_{\rm p} (t-t_1)}
\end{equation}
is an auxiliary impulse response of the output field amplitude $ \hat{\rm y}(t)$ on the input one $ \hat{\rm z}(t)$ in the rotating frame, i.e.~excluding the time evolution of the phase.
Explicitly, the function is
\begin{equation}\label{LN2}
L_{\rm s}(t,t_1)= \sum\limits_{n=0}^{\infty}B_n\exp\left[i\varphi_n(t)\right] \delta (t_1-t +n\tau)
\end{equation}
with amplitudes $B_n$ and phases $\varphi_n(t)$ defined in \eqref{phsh}.

A more detailed derivation of the impulse response is presented in Appendix \ref{app:GEO}.
\begin{figure}
  \includegraphics [angle=0,width=0.46\textwidth]{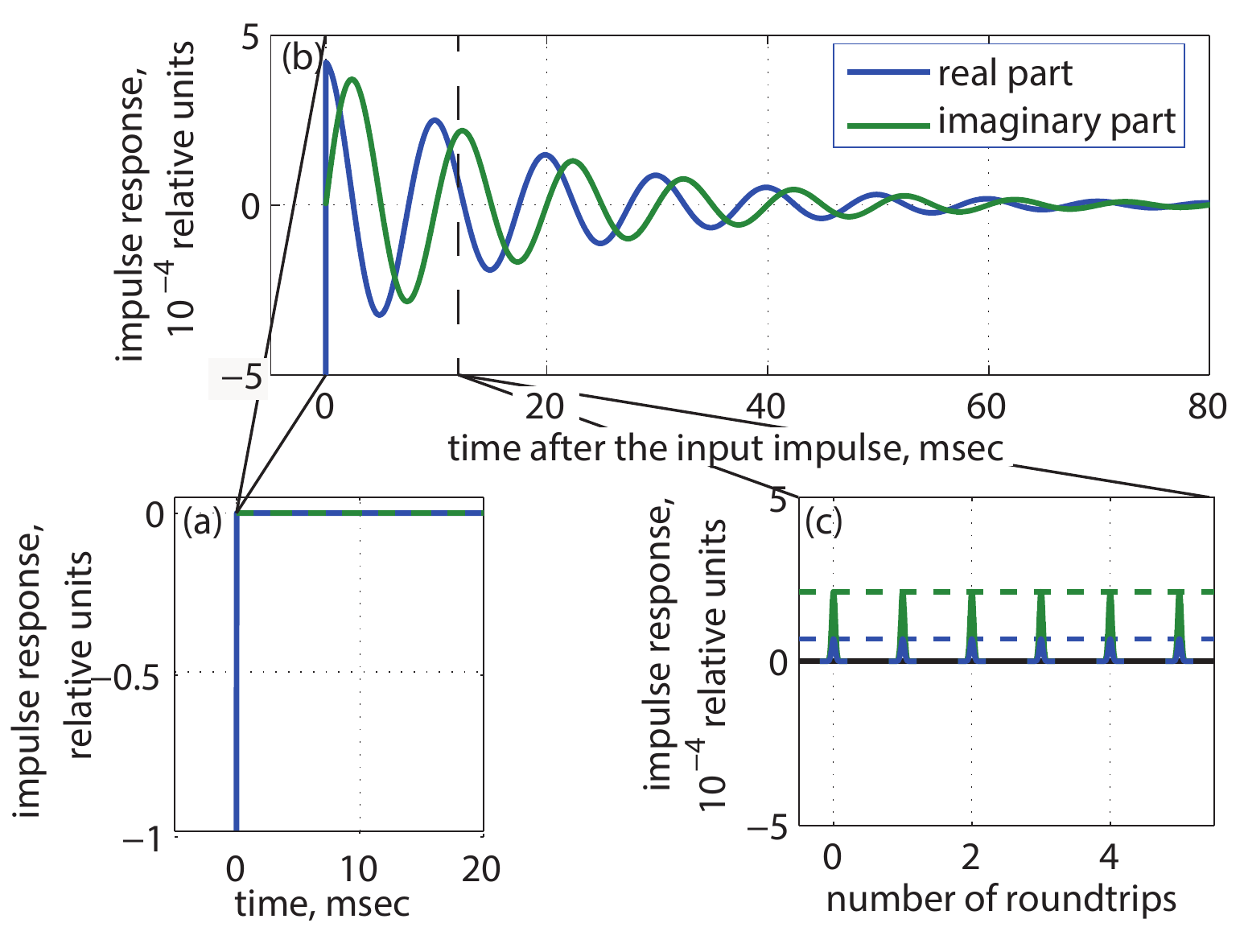}
  \caption{Two orthogonal quadratures of the auxiliary impulse response (real and imaginary parts) to the vacuum quantum oscillations, injected from the dark port: (a) on the large amplitude scale, depicting the direct reflection of the input pulse from the SRM; (b) on the small amplitude scale, representing the output envelope from the the oscillations and decay (in the rotated frame) of the amplitude pulse as it propagating inside the SRC and (c) on the short time scale, picturing the round-trip time and the discrete nature of the impulse function.}\label{fig:impulse2}
\end{figure}

The plot of the two quadratures of the auxiliary impulse response is presented in Fig.~\ref{fig:impulse2}. The deltalike impulse, sent to the dark port $\hat{z}(t)$, is reflected back from the SRM almost completely ($R_{\rm s}\approx 1$), as can be seen in Fig.~\ref{fig:impulse2}(a). Only a tiny fraction of the input pulse is injected into the detector and does the round trips the way it was thoroughly described in Sec. \eqref{detRespGW}. For every round trip, when the pulse reaches the SRM, a part of it goes to the output [see Fig.~\ref{fig:impulse2}(c)]. On a larger time-scale the pulses in both quadratures form the oscillating and decaying envelopes shown in Fig.~\ref{fig:impulse2}(b).

The transfer function $R_{\rm s}(\Omega)$ \eqref{TransFuncDP} for the stationary regime can be obtained from Eq. \eqref{LN2}. It is consistent with the results obtained by the conventional calculations in the frequency domain  \cite{rakhmanov00, Harms03, danilishin12, simakov14}. In the case of the ideal end mirrors $R_{\rm f} = -1$, the transfer function turns to the expression \eqref{TransFuncDPid}, the modulus of which equals 1.

The auxiliary impulse response \eqref{LN2} with the help of Eq. \eqref{vacautocor} and with some simplifications leads us to the autocorrelation function of the output shot noise,
\begin{multline}\label{outcorr8}
 B_{\rm \eta}(t_1,t_2) = C_{\rm z}\sum\limits_{n=-\infty}^{\infty}D_n\cos(\varphi_{n+1}(t_1))  \delta(t_1-t_2-n\tau)+\\
+C_{\rm z}\delta(t_1-t_2),
\end{multline}
the amplitudes of the correlations $D_n$ of which are presented in \eqref{outcorrcoef}. This is a correlation of white noise reflected from the cavity. The bigger part of the wave reflects directly, remaining delta correlated, and the part transmitted inside makes a sequence of echoes every round trip, each of which is also correlated with the directly reflected wave and with other echoes. Because of the properties of white noise, neither echoes, nor directly reflected light correlates with the pieces of wave between the echoes.

As it follows from the expression for coefficients \eqref{outcorrcoef}, the autocorrelation function for the detector with ideally reflecting end mirrors $\left|R_{\rm f}\right| =-1$ keeps only one nonzero summand $C_{\rm z}\delta(t_1-t_2)$, meaning the detected shot noise is white. Despite the nontrivial transformation of the electromagnetic field of the quantum oscillations inside the detector \eqref{LN2}, its statistics remains the same. 

The delta-correlated statistics in this case is also consistent with the conventional description of the stationary interferometers in the frequency domain. As it was mentioned before, in the case of the ideally reflecting end mirrors the modulus of the transfer function \eqref{TransFuncDPid} equals 1. The spectral density of the shot noise in this case is proportional to the squared modulus of the transfer function, and therefore constant, also meaning the noise is white. 
\begin{figure}
  \includegraphics [angle=0,width=\columnwidth]{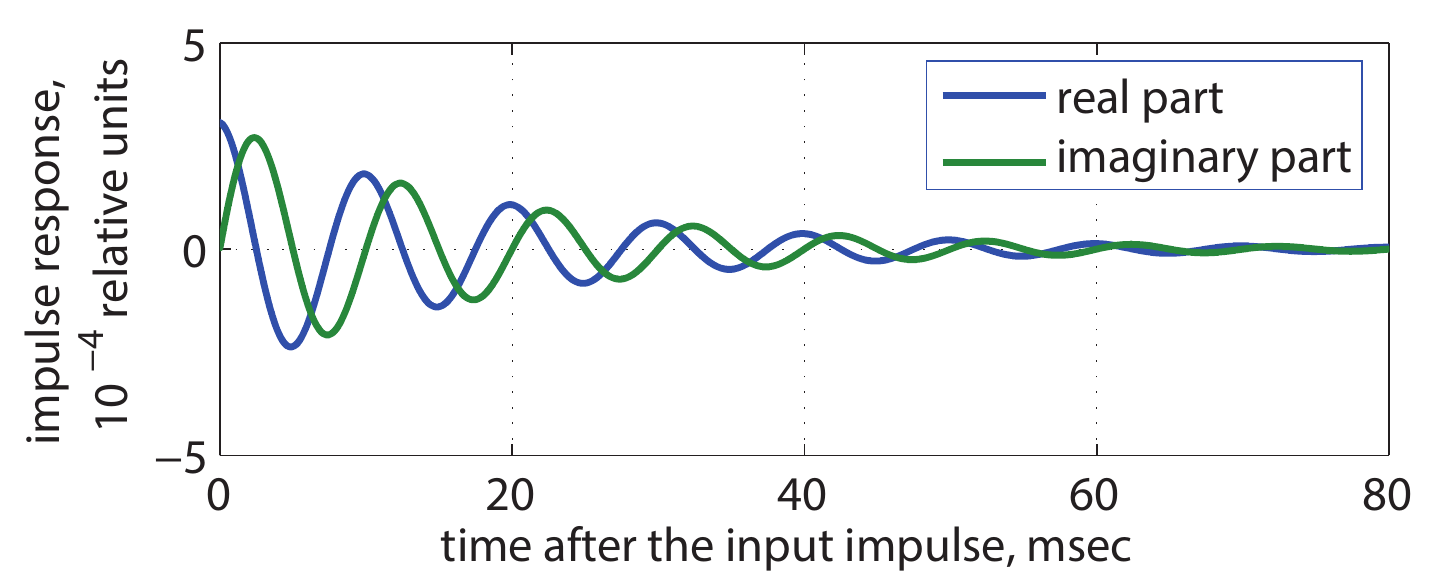}
  \caption{Two quadratures of the auxiliary impulse response (real and imaginary parts), depicted on the decay time scale, to the vacuum quantum oscillations, injected from an equivalent end mirror. The difference between the east and the north end mirrors is insignificant on the plot scale.}\label{fig:impulse3}
\end{figure}

\subsection{Influence of losses}
The optical losses of the laser field inside a cavity, caused by the scattering into higher-order modes, reflection from the antireflective coating of the beam splitter, and the absorptions in all optical elements, decreases the effective reflectivity of the equivalent mirror $R_{\rm f}<1$ and therefore modifies the statistics of the corresponding output noise in \eqref{outcorr8}. However, the losses cause additional noise due to the fluctuation-dissipation theorem \cite{welton51}, which can be equivalently considered as injections of the ground-state vacuum quantum oscillations through the equivalent mirrors \cite{KLMTV} with a transmittance equal to the optical losses in the arms ($A_{\rm e}$ and $A_{\rm n}$ in Fig.~\ref{fig:geo1}). The corresponding impulse responses read in a similar manner to \eqref{LN2}:
\begin{subequations}\label{Aimpulse2d}
\begin{multline}\label{Aimpulsen2d}
L_{\rm s,n\rightarrow c} (t,t_1)= \sum\limits_{k=0}^{\infty}B_{{\rm n}k} \exp[i\varphi_{k+1}(t)]\times \\
\times \delta\left(t_1-t+\frac{\tau}{2}+k\tau\right),
\end{multline}
\begin{multline}\label{Aimpulsee2d}
L_{\rm s,e\rightarrow c}(t, t_1) = \sum\limits_{k=0}^{\infty}B_{{\rm e}k}\exp[i\varphi_{k+1}(t)]\times \\
\times   \delta\left(t_1-t+\frac{\tau}{2}+k\tau\right).
\end{multline}
\end{subequations}
The coefficients $B_{{\rm n}k}$ and $B_{{\rm e}k}$ are explicitly presented in \eqref{Aimpulsecoef} and the phase shifts $\varphi_n(t)$ are the same as in the previous section and defined in \eqref{phsh}.

The plots of the impulse responses are depicted in Fig.~\ref{fig:impulse3}.

The autocorrelation function of their noise on the photodetector caused by both of these injections is
\begin{multline}\label{outcorrF1}
B_{\rm \eta}(t_1,t_2) = -C_{\rm z}\sum\limits_{n=-\infty}^{\infty}D_n\cos(\varphi_{n+1}(t_1))  \delta(t_1-t_2-n\tau).
\end{multline}

As it is easy to see, the total output shot noise of the nonstationary detector with the moving SRM, including the injections of the ground-state oscillation from the dark port (\ref{outcorr8}) and from the optical losses (\ref{outcorrF1}), is white:
\begin{equation}\label{outcorrtot}
B_\eta^{\rm tot}(t_1, t_2) = C_{\rm z} \delta(t_1-t_2).
\end{equation} 

This result is expected because the autocorrelation function of the output shot noise defines the state of the output electromagnetic oscillations at the output. There are no generators of photons inside the detector contributing to these oscillations, even though we move the SRM. Therefore the state of the output shot noise should also be ground and should have the same statistics as the input electromagnetic oscillations.

The SNR at the output with respect to the white shot noise is determined by Eq. \eqref{SNRhom_main}.

\section{Dynamical behavior of a nonstationary detector}\label{DynBeh}
The important goal of the investigations presented in this paper is the study of the dynamical behavior of the detector, resulting from the nonstationarity caused by the fast movement of SRM. The analysis is based on the model for this regime presented in Sec. \ref{detRespGW} with the impulse response.

Dynamical tuning assumes that we are detecting a chirp GW signal. Because of their sinusoidal shape, these signals can be presented in form
\begin{equation}\label{inSin}
x_{\rm d}(t) =X(t)\cos \zeta(t),
\end{equation} 
where $X(t)$ and $\zeta(t)$ are the time-dependent amplitude and phase, respectively. The latter is related to the time-dependent frequency of the signal: $\zeta(t) = \int\limits_{t_0}^{t}\Omega(t_1)dt_1$, where $\Omega(t) = 2 \pi f(t)$ is an angular frequency.

\subsection{Quasistationary dynamical tuning}
A quasistationary approximation is the first and the simplest approach to describe a dynamically tuned gravitational wave detector. It was presented and described in the pioneer paper about dynamical tuning by Meers {\it et al.} \cite{meers93}. According to this approximation the parameters of the gravitational waves, namely the amplitude and the frequency of the gravitational wave, and also the detuning of the signal-recycling cavity, change slowly enough, so the fields inside the cavity and the current at the homodyne detector reach their stationary values, and the transients are negligible \eqref{domQuasyStat}. 

In Appendix \ref{app:quasstat} it is shown that dynamical tuning in quasistationary approximation amplifies a GW uniformly, without deformations. So the output signal in a quasistationary case equals a multiple of the GW,
\begin{equation}\label{outQuas}
I_{\rm y}(t) = C_{\rm qs} x_{\rm d}(t),
\end{equation}
with the coefficient $C_{\rm qs}$ presented in \eqref{Cqs}. 

The {\it dynamical processes} (or {\it dynamical behavior}) studied in this paper are defined as the difference between the nonstationary time-domain and the quasistationary models. It arises outside the domain of applicability of a quasistationary approach \eqref{domQuasyStat}  [see e.g.~Figs.~\ref{fig:gw}(c) and \ref{fig:snr}].

\subsection{Resonant tracking of the sinusoidal signal} 
During dynamical tuning, the cavity is resonant to only one of the sidebands caused by a GW, while the other one is suppressed. The tuning of the sideband takes place, when the additional phase shift the GW gets during one round trip is canceled by the corresponding displacement of the SRM from the laser resonance position. Generally speaking, this condition is defined within one round trip and therefore we can express it mathematically for the nonstationary detector with moving SRM:
\begin{equation}\label{resonantCond}
2\pi \delta_{\rm f}(t)\tau =\zeta(t+\tau/2)-\zeta(t-\tau/2)\approx 2\pi f(t) \tau.
\end{equation}
It is easy to prove this resonance condition by the substitution of the resonant condition \eqref{resonantCond} to the impulse response \eqref{impulses2pd} and applying it to detect the GW \eqref{inSin}. The dynamical tuning detection, following this resonance condition, is referred to as {\it resonant tracking} of the signal. The similar task of the {\it dynamic resonance} of a Fabry--Perot cavity to the perturbations of the laser phase inside it is considered in details in Ref. \cite{rakhmanov02}.

\subsection{Transient features and virtually stationary detection}
\begin{figure}
  \includegraphics [angle=0,width=\columnwidth]{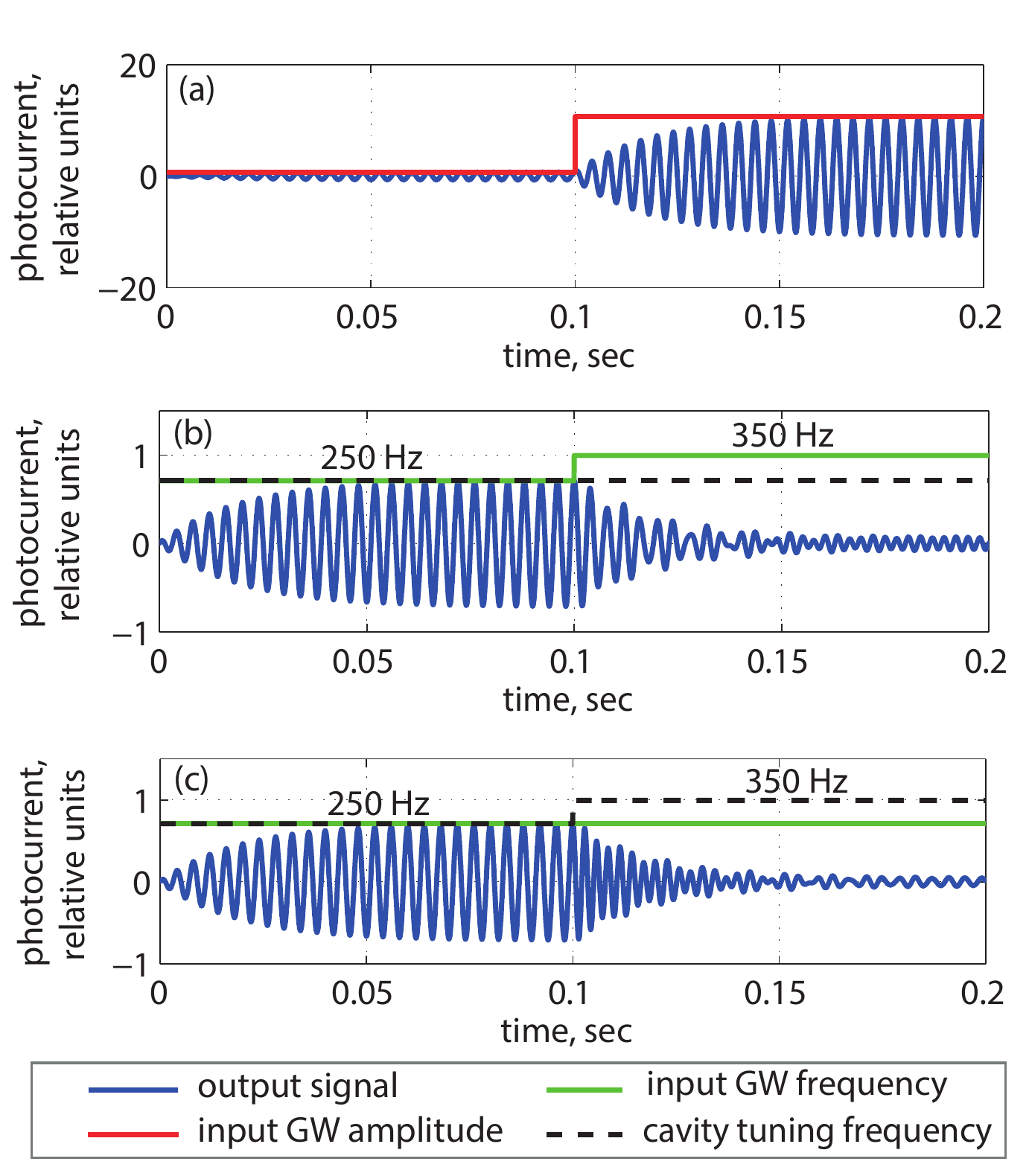}
  \caption{The typical transients of the considered detector on the stepwise change of (a) $X(t)$, (b) $f(t)$, and (c) $\delta_{\rm f}(t)$}.\label{fig:trans1}
\end{figure}
From the mathematical point of view, in the quasistationary approximation described above, the system switches from one stationary state into another one instantaneously. In real systems, there are often finite transient processes between two stationary cases, caused by inner physical phenomena, making the behavior of these systems purely nonstationary. These dynamical transient effects of the detector make the essential difference between the known quasistationary approximation and the new time-domain model presented in this paper. 

The major transient features of the nonstationary detector become apparent already in the plot of its linear response, depicted in Fig.~\ref{fig:imp}. The decaying oscillations of the envelope (created by the beating of the detector sideband with a local oscillator) represent, respectively, the detuning (which can also be time dependent), and the relaxation time of the SRC, while the delays between impulses equal the round-trip time.

The numerical simulation of the output photo current \eqref{linres} using the impulse response \eqref{impulses2pd} allows us to study the dynamical transient effects from the stepwise change of the parameters. In this paper, we focus on the amplitude $X(t)$, the signal frequency $f(t)$, and the detuning of the SRC $\delta_{\rm f}(t)$, since only they are changing during dynamical tuning. Strictly speaking, only the change of $\delta_{\rm f}(t)$ makes the detection nonstationary. The response on $X(t)$ and $f(t)$ alone is a response of a stationary detector. However, when the detector is nonstationary, the changes of these two values introduce additional nonstationary effects, as it is shown below. The numerically simulated transients from the stepwise changes are shown in Fig.\ref{fig:trans1}. In all three cases the signal starts with the same set of parameters $X(t)$, $f(t) = \delta_{\rm f}(t)= 250 {\rm Hz}$, followed by a stepwise change of one of them.

The first typical feature of all three transients is the length of the relaxation processes, depicted in all three transients. These relaxations have the same duration as the impulse response (see Fig.~\ref{fig:imp}).

Let us consider the transients caused by the changes of a frequency: either of the GW signal in Fig.~\ref{fig:trans1}(b) or of the SRC detuning in Fig.~\ref{fig:trans1}(c). In the transient caused by the signal frequency change we can see the decay of oscillations at the initial signal frequency. The new stationary oscillations, arising during the transient, occur at the new frequency of the GW. In the other transient, caused by the change of the SRC detuning, the decaying oscillations immediately occur with the new detuning frequency. The new stationary oscillations arise here with the GW frequency. Generally speaking, the energy stored before the parameter shift decays at the new SRC detuning frequency, while the new oscillations arise at the new GW frequency.
\begin{figure}
  \includegraphics [angle=0,width=\columnwidth]{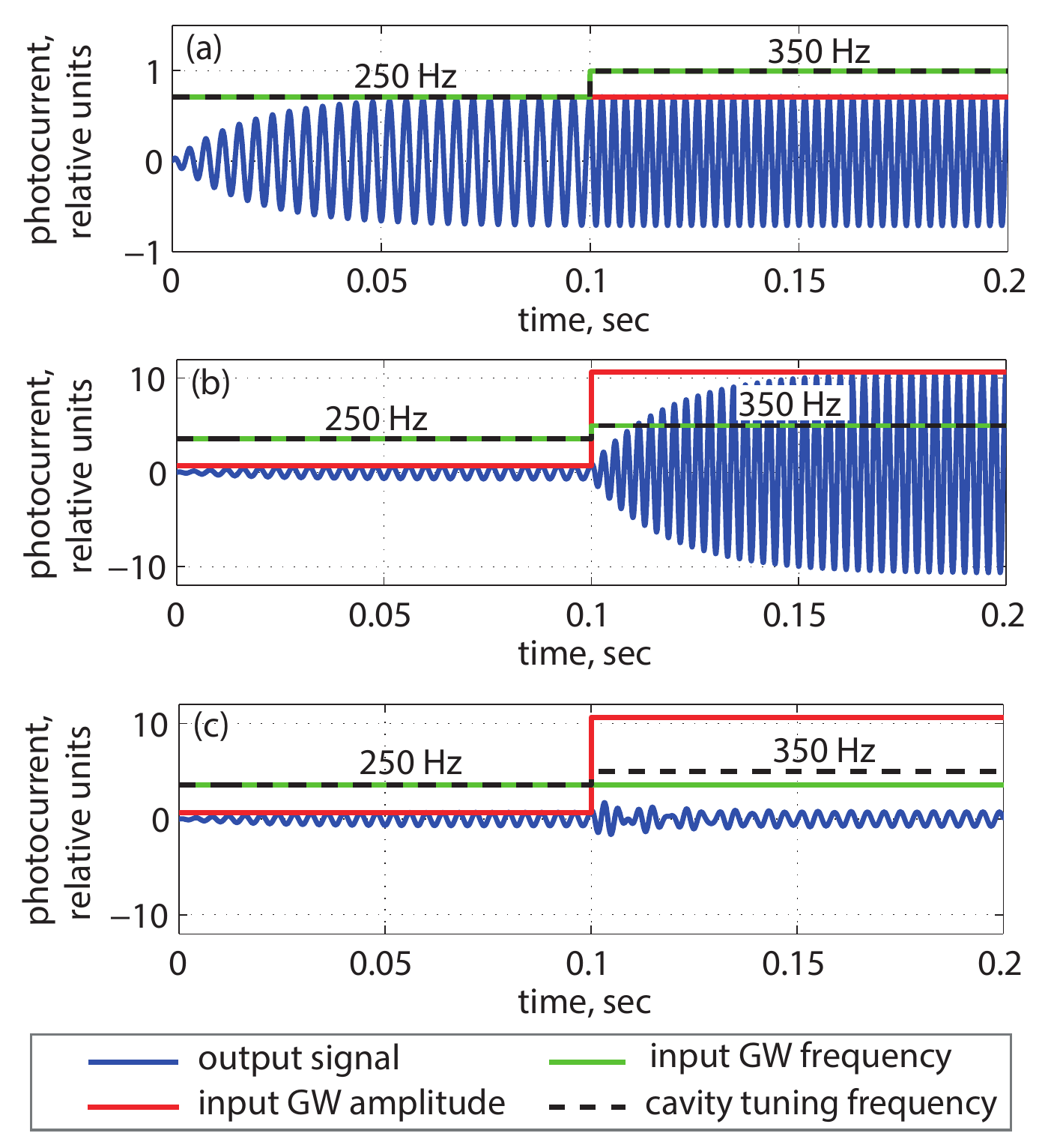}
  \caption{The transients of the considered detector on the combinations of stepwise changes of (a) $f(t)$ and $\delta_{\rm f}$, (b) $X(t)$, $f(t)$ and $\delta_{\rm f}(t)$, (c)$X(t)$ and $\delta_{\rm f}(t)$.}\label{fig:trans2}
\end{figure}

All the explained conclusions are easy to obtain mathematically by the substitution of a corresponding stepwise changing parameter into the impulse response equation \eqref{impulses2pd}. It was performed explicitly in Re. \cite{simakov14}.

The remarkable consequence of the described processes becomes apparent as we numerically simulate the resonant tracking \eqref{resonantCond} to the step change of the GW frequency, i.e.~when both frequencies of the GW and of the SRC detuning are changing synchronously as it is depicted in the Fig.~\ref{fig:trans2}(a). The decay at the SRC frequency is compensated here by arising at the same GW frequency. As a result, the perturbations caused by these two transients are canceled, so the frequency of the output signal switches instantaneously from one value to the other without any relaxation process.

Since any arbitrary change of frequency may be expressed in terms of single step transients, the following can be deduced. During the resonant tracking of a chirp GW with constant amplitude, the output current will have its stationary values everywhere, without transition effects, although the detector could be rather nonstationary. This condition we call virtually stationary. Consequently, only a change of GW amplitude causes transients.

\subsection{Transformation of the signal envelope during resonant tracking}
The dynamical effects become crucial during the late stages of resonant tracking. 

\begin{figure}
  \includegraphics [angle=0,width=\columnwidth]{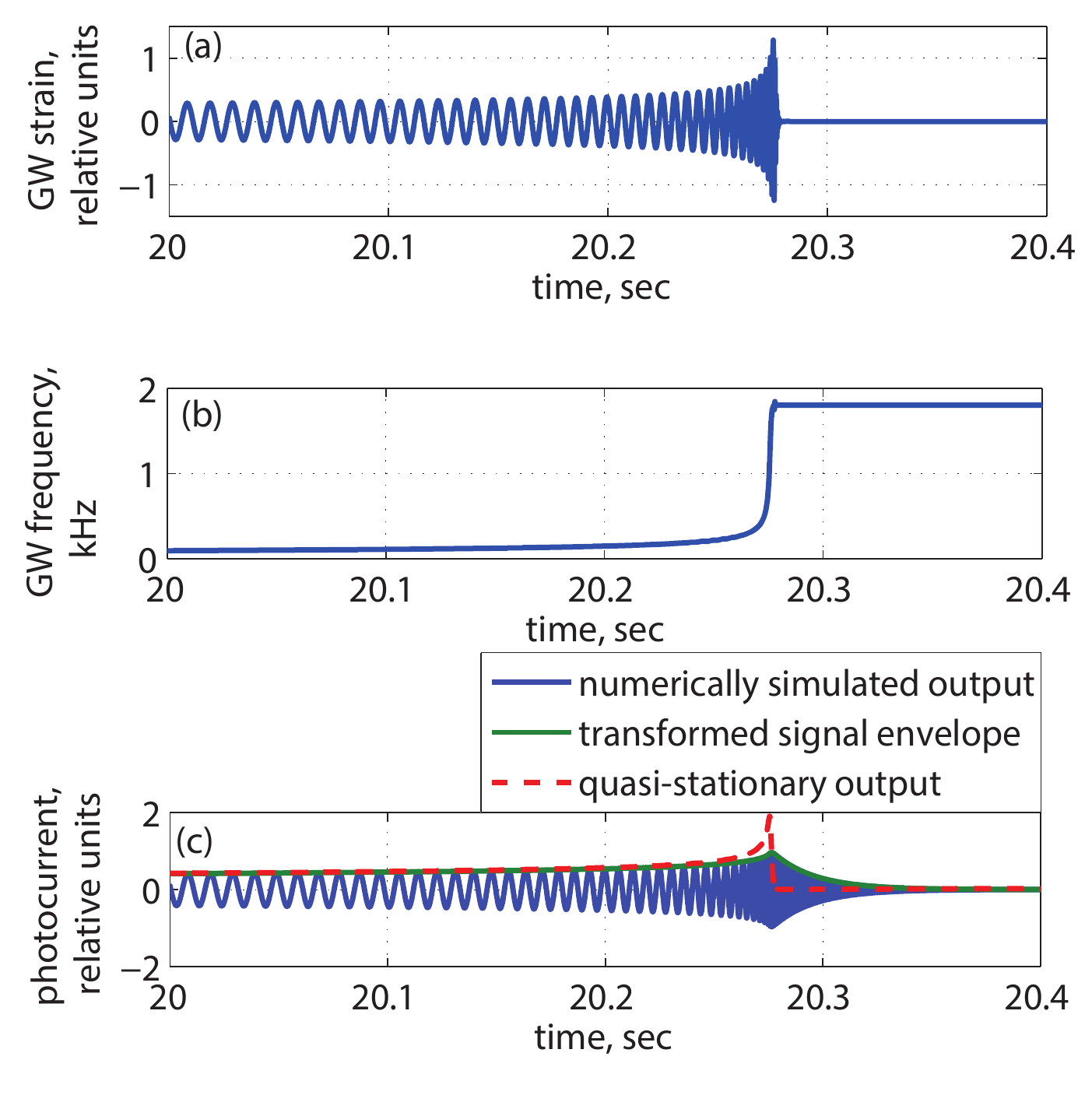}
  \caption{(a) The gravitational wave signal from a 5+5 Solar mass spinless black hole binary. (b) The instantaneous frequency of this signal. (c) The output for the resonantly tracked detection of this signal, calculated using three models: numerical simulation, a transformed envelope, and a quasistationary approximation.}\label{fig:gw}
\end{figure}

Let us calculate the response of the detector to a chirp GW signal \eqref{inSin}. The approach explained here is shown in more mathematical details in Appendix  \ref{app:restrack}. An example of such a signal from a coalescing binary and of its frequency behavior are shown in Figs.~\ref{fig:gw}(a) and \ref{fig:gw}(b). Generally, the approach described in this subsection is valid for signals with arbitrary changes of phase and amplitude. The phase of the chirp defines the required motion of the SRM \eqref{resonantCond}. The linear response, defined by this SRM motion via Eq. \eqref{impulses2pd}, helps to calculate the output \eqref{linres}.

The resulting expression consists of an infinite series, each summand of which includes two cosines, one with the phase of the GW from Eq. \eqref{inSin} and one for the phase of the cavity detuning from \eqref{impulses2pd}. We can expand both cosines into complex exponents and choose only the resonant sideband terms from their product, assuming the nonresonant fields, being summed up with homogeneously distributed phases, to be insignificant. The amplitude of the gravitational wave $X(t)$ in the same expression can be represented in the Fourier domain as $X(\Omega)$, which turns time delays into complex exponents. Subsequent reducing of the geometric series in the obtained formula, brings us to the result for the output photocurrent,
\begin{equation}\label{outSin}
I_{\rm y}(t) = Y(t)\cos\zeta(t-\tau/2),
\end{equation} 
with the time-dependent amplitude $Y(t)$ and the same phase behavior as the input gravitational wave.
It appears that the mathematical expression for the output amplitude $Y(t)$ may be expressed from the GW amplitude $X(t)$
where the Fourier transform of $Y(t)$ reads
\begin{equation}\label{ampTrans}
Y(\Omega) = R(\Omega)X(\Omega).
\end{equation}
The transfer function $R(\Omega)$, the explicit expression of which is presented in \eqref{transFunc}, is an Airy function for the equivalent Fabry--Perot cavity. Its frequency half-bandwidth $\gamma$, depending on the optical parameters of SRM and optical losses with \eqref{freqBW}, is 8.3\,Hz.

The phase and the frequency behavior of the output signal repeats those of the input GW, while the amplitude at the output is smoothed with respect to the amplitude of the GW signal. In other words, during the resonant tracking the output signal may be obtained from the GW signal by low-pass filtering its amplitude.

Obviously, when the amplitude of a gravitational wave $X(t)$ changes slowly enough, i.e.,~its typical frequency components are small with respect to the detector half-bandwidth
\begin{equation}
\Omega \ll \gamma,
\end{equation} 
the output signal will have both amplitude and frequency repeating those of the gravitational wave. So, under this condition, a virtually stationary detection is performed.

In Fig.~\ref{fig:gw}(c), three different results are presented for the output signal from the resonant tracking of the chirp signal, depicted in Fig.~\ref{fig:gw}(a): (blue) the one, simulated numerically using the linear response from (\ref{linres}, \ref{impulses2pd}); (green) its envelope, calculated using transfer function $R(\Omega)$ \eqref{ampTrans}; and (dashed red) the output, calculated using a mathematical model for the quasistationary approximation \eqref{outQuas}. The transformation of the envelope, presented in this subsection, is obtained by neglecting the components of the nonresonant sideband. The comparison of the results of the numerical simulation and of the envelope transformation confirms the negligible influence of these components on the output. 

Comparison of the green and red dashed lines in Fig.~\ref{fig:gw}(c) shows the difference between the old quasistationary model and of the new time-domain models of dynamical tuning. They agree at the earlier stages of the signal, but diverge during the later stages due to  dynamical effects.

\subsection{Deconvolution of the signal}
With the previous subsection, one can find the obvious way of restoring of the GW signal after the resonantly tracked detection. The division of the output envelope by the transfer function \eqref{ampTrans} gives the Fourier transform of the envelope of the input GW. However, there are always errors in the tuning of the SRC to the signal, reducing the applicability of this restoring.

However, if we could know the exact (though not resonant) motion of the SRM, it would be possible to find the inverse impulse response from \eqref{impulses2pd} and \eqref{2pddet}. The output signal at the instance $t+\tau/2$ of time carries information about the end-mirror displacement in the infinite number of the previous moments of time $t-n\tau$, with natural $n$. To single out the information about only one displacement, the others should be subtracted, which is possible, using the previous output signals, which contain the influence only of the previous, with respect to the required displacements. The more detailed explanation is described in Appendix \ref{ssec:GEO_dif_inv}. The explicit expression for it reads
\begin{equation}\label{invImpFunc}
L_{\rm c\rightarrow s}(t,t_1) =\sum\limits_{n=0}^{\infty}\tilde{A_n}(t)\delta(t_1 - t-\tau/2+n\tau).
\end{equation}
The factors $\tilde{A}_n(t)$ defined explicitly in \eqref{invImpFuncDef} express the filtering out of the required component of the output signal. The inverse impulse response allows  by definition the deconvolution of the gravitational wave shape, using the signal on the photo diode and the known motion law of the SRM, without restrictions on the tuning or on a GW.

The following equation proves that the eigenbasis of both direct and inverse impulse response transformations is full, so theoretically no information about the GW signal is lost during the resonant tracking:
\begin{equation}\label{invMatDet}
\int\limits_{-\infty}^{\infty}L_{\rm c\rightarrow s}(t,t_1)L_{\rm s\rightarrow c}(t_1,t_1') = \delta(t-t_1').
\end{equation}

\section{Sensitivity gain from dynamical tuning}\label{simRes}
\begin{figure}
  \includegraphics [angle=0,width=\columnwidth]{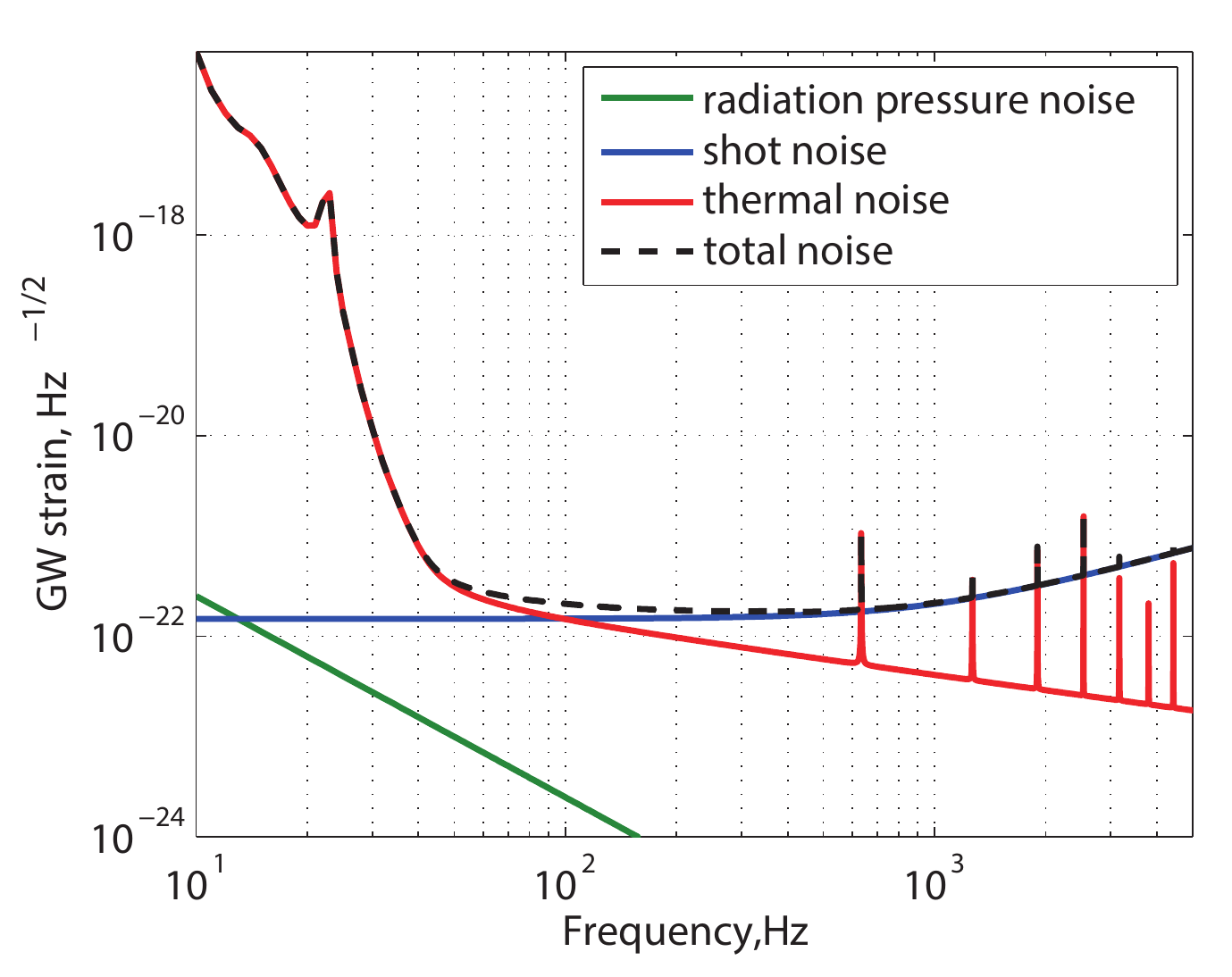}
  \caption{The theoretical noise budget of the current GEO\,600 configuration.}\label{fig:noises}
\end{figure}


The main goal of any detector development is increasing the sensitivity. In this section we study what sensitivity gain could be achieved by the implementation of the dynamical tuning. As a reference we use the current sensitivity of GEO\,600 operating in the stationary broadband regime. The output signal and noise are largely dependent on the parameters of GEO\,600. They are presented for both regimes in Tables \ref{tab:GEOconst} and \ref{tab:GEOvar}.
\begin{table*}
\caption{Unchanged GEO\,600 parameters.}
\centering
\begin{tabular}{c l c}
\hline
 Symbol        &quantity         & Current configuration         \\
               &                 & value                         \\
\hline
\hline
    $A_{\rm e}^2$             &Equivalent power transmission on  the east mirror &    	 $450$ ppm   $(10^{-6})$						   \\
  &   (losses at the mirrors + scattering on the  beam splitter)  &    \\              
\\
$A_{\rm n}^2$   &Equivalent power transmission on the east mirror  &    			$390$ ppm		     \\
               &(losses at the mirrors)&                         \\

\\
L              & Effective   length of the arm      &$1200$ m                       \\
\\
$W_{\rm e}$          &Power falling on the beam splitter  (at point E in Fig.\ref{fig:geo1})  &2.12 kW                            \\
\hline 
\end{tabular}
\label{tab:GEOconst}
\end{table*}

\begin{table*}
\caption{GEO parameters, modified for the dynamical tuning.}
\centering
\begin{tabular}{c l c c}
\hline
 Symbol        &quantity         & Current configuration       &Value for dynamical   \\
               &                 & value                       &tuning configuration  \\
\hline
\hline
$T_{\rm s}^2$          &Power transmission  on the SRM  &              0.1             &      420 ppm          \\
               \\                    
$\delta$       &Frequency detuning of signal-recycling cavity      &$0$\,Hz                 & Resonant tracking  \\                       
\hline

\end{tabular}
\label{tab:GEOvar}
\end{table*}

The typical value describing the sensitivity is a signal-to-noise-ratio that can be derived from both Wiener filtering \cite{thorne87} and the Neyman--Pearson criteria (see Appendix \ref{sigDet}). For GW detectors, it is usually used in the frequency domain for stationary noise [see e.g.~the SNR for displacement noise \eqref{SNR_disp}]. However, the concept of SNR is also applicable for the regimes with non-stationary noise, assuming the noise is Gaussian (which is a good approximation for GEO\,600 after the vetoing of glitches).

The noise of the detector may be divided into three parts: shot noise, radiation pressure noise and displacement noise. The theoretical curves of these noises for GEO\,600 are pictured in Fig.~\ref{fig:noises} \cite{geonoise, har}. We assume that the real noise will be reduced to the theoretical predictions, and we consider only them for the analysis.  The radiation pressure noise is negligible in the frequency band of our interest.

The computation of the SNR in the case of signal-to-shot-and-displacement-noise-ratio is quite a complex task in the nonstationary case. However the consideration of the displacement or shot-noise-limited nonstationary detector gives already the boundaries that are realistic estimations for the sensitivity gain from dynamical tuning.

Chirp signals, for which we want to increase the sensitivity by dynamical tuning, are modeled by using hybrid models \cite{santamaria10, ajith11, kamaretsos12, kamaretsos12b} for an arbitrary set of masses and spins of the binary elements. For convenience only one group of signals is analyzed: spinless binaries with equal masses and total mass ranging from 3 to 10 solar masses. The significant benefits from the dynamical tuning arise at the very last stages of the chirp, when the typical signal frequencies are in the shot noise-limited frequency band of the detector. The rate of frequency change also becomes high at this stage, causing the nonstationary effects, described in Sec. \ref{DynBeh}. To consider the important part of a chirp signal and to avoid the influence of radiation pressure noise, we consider for each signal only the segment starting with the instantaneous frequency 200\,Hz, as it is shown, for example, in Fig.~\ref{fig:gw}.

According to Sec. \ref{detRespSN}, the shot noise at the output of a nonstationary detector is white. Therefore, it is convenient to calculate the sensitivity for the shot-noise-limited detector at the photocurrent, too. The output signals are simulated numerically according to the algorithm from Appendix \ref{sigAlg}, based on the time-domain model, described in Sec. \ref{detRespGW} with help of the linear response (\ref{impulses2pd}, \ref{2pddet}).

 \subsection{Shot-noise-limited dynamically tuned detector vs shot-noise-limited reference detector}\label{subsec:snsn}
 
  \begin{figure}
  	\includegraphics [angle=0,width=\columnwidth]{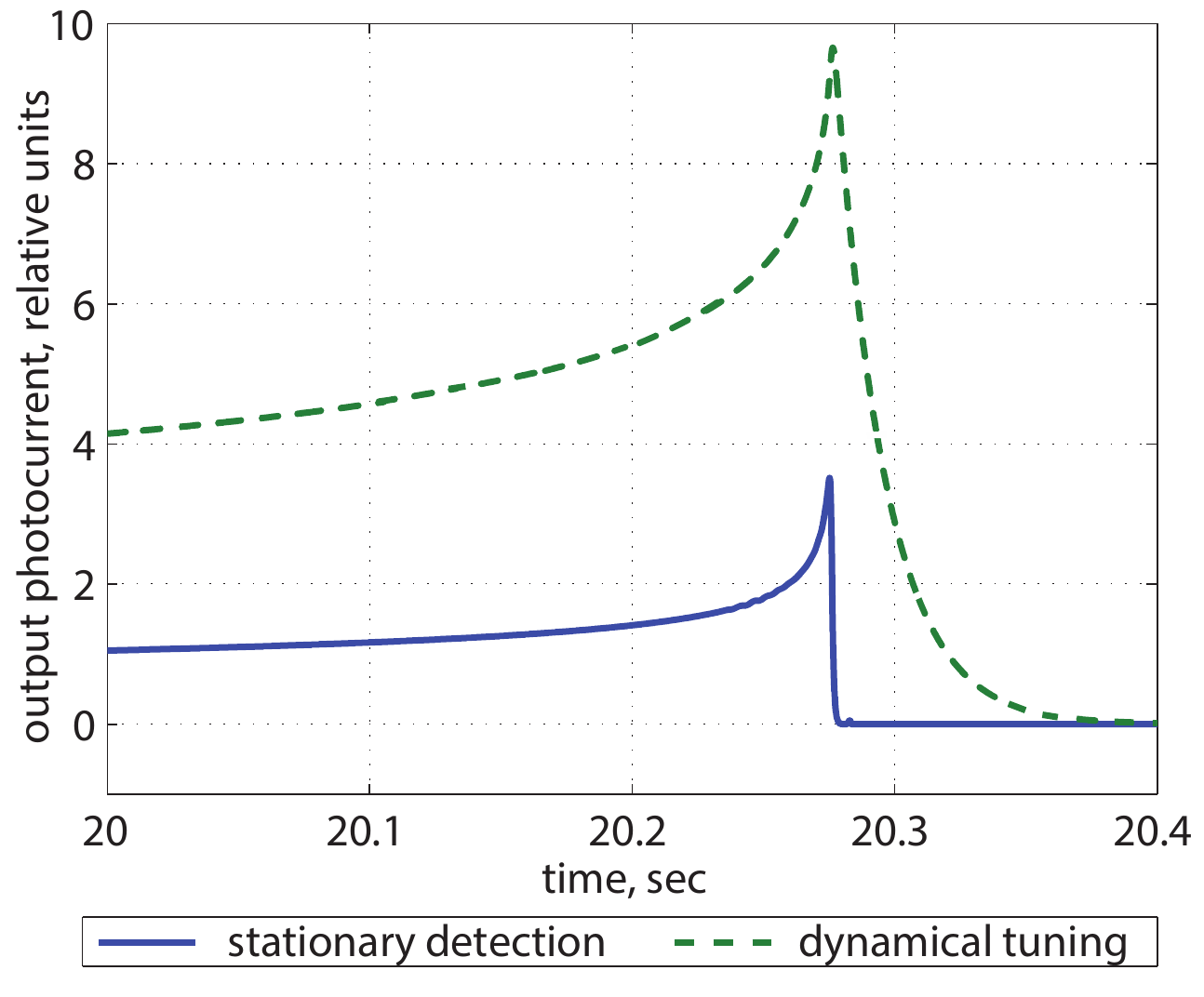}
  	\caption{Comparison of the output signal envelopes from the dynamical tuning with respect to the reference stationary detection. The detected GW is from a 5+5 Solar mass spinless black hole binary (see Fig.~\ref{fig:gw})} \label{fig:compOut}
  \end{figure}
 
 \begin{figure}
   \includegraphics [angle=0,width=\columnwidth]{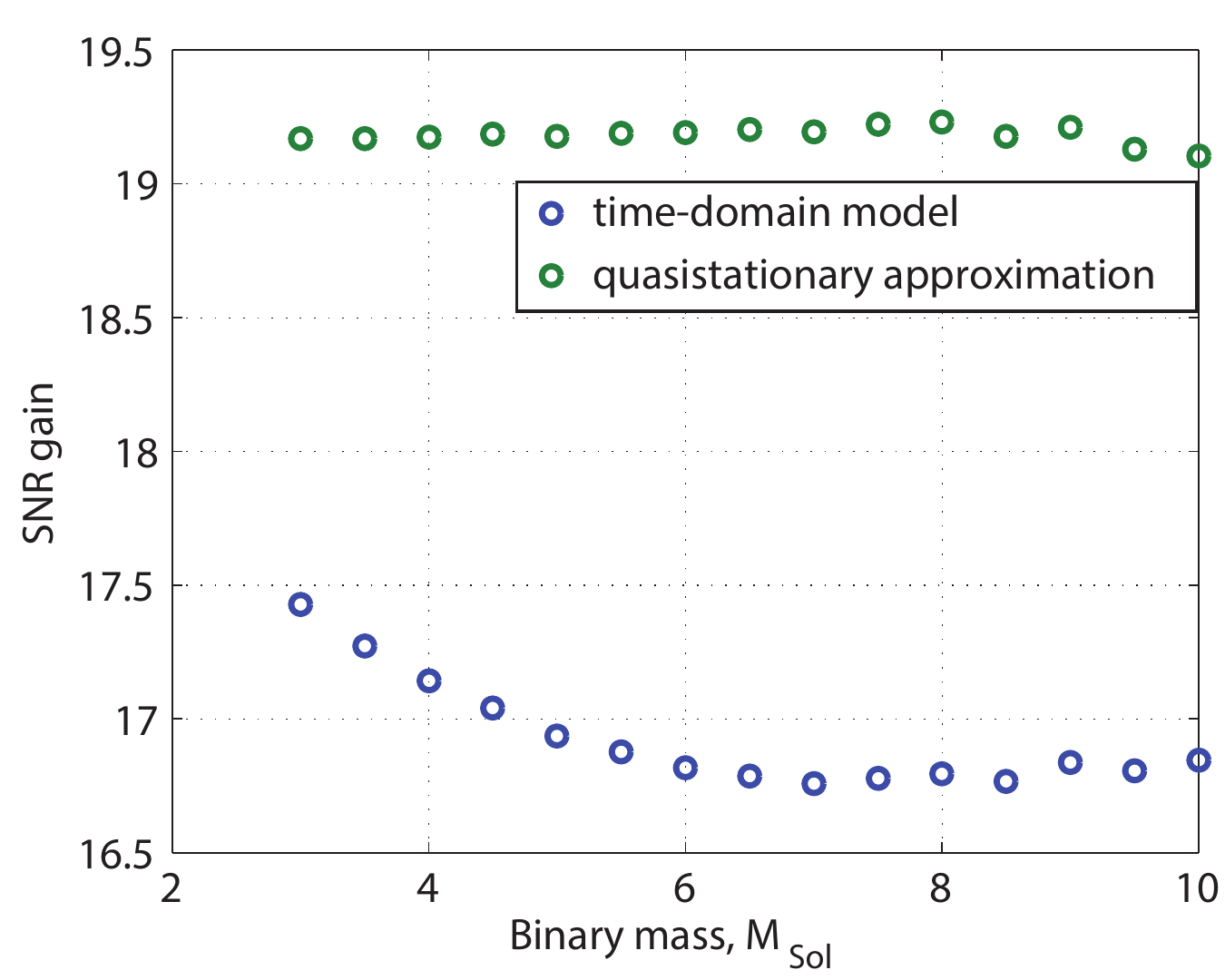}
   \caption{The SNR gain from implementation of dynamical tuning into a shot-noise-limited detector that worked in a stationary broadband regime. The influence of the nonstationary dynamical effects on it becomes apparent in comparison with the depicted result of a quasistationary approximation.}\label{fig:snr}
 \end{figure}
 The sensitivity with respect to the displacement noise is the same for the reference detectors and for the detector with dynamical tuning, because each of them transforms the similar components of both GW and displacement noise in the same way. Thus, the benefits from dynamical tuning arise only with respect to shot noise. For this reason we first compare the two detectors limited by shot noise. It also allows us to study the pure influence of the dynamical tuning and the dynamical behavior on the sensitivity. 

 The ground-state shot noise on the photodiode remains delta correlated and has the same intensity independent from the parameters of the SRM, namely, its motion during the detection and its transmittance (see Sec. \ref{detRespSN}). Therefore, once we are limited by shot noise, only the transformation of the signal by both regimes defines the sensitivity gain. This allows us to study the influence of the dynamical effects, described in Sec.~\ref{DynBeh}. Thus, even though GEO\,600 is currently operating with squeezed shot noise, we consider here the ground-state shot noise for the referent detector.

 Both output signals, in the dynamically tuned and in the reference detectors, are obtained by numerical simulations, based on the time-domain algorithm. The detection of the stationary detector can also be simulated in the frequency domain using the transfer function. The calculations in the frequency domain and in the time domain differ, however, only by a Fourier transform, and apart from this are equivalent. The example of the output for a dynamically tuned chirp GW signal is presented in Fig.~\ref{fig:gw}(c) as a blue line with a green envelope. The source for the GW in this example is a compact binary, consisting of two black holes with five Solar masses each. Both output signals for this input, detected by dynamical tuning and by the stationary detector, are presented in Fig.~\ref{fig:compOut}.

 Once the photocurrent signal is calculated, the SNR for it in the shot-noise-limited case may be calculated according to the following formula \eqref{SNRhom}, derived in Appendix \ref{sigDet}:
 \begin{equation}\label{SNRhom_main}
 d^2 = \frac{1}{C_{\rm z}}\int\limits_{0}^{T} s^2(t)dt.
 \end{equation}

 There are two equally used values typically called SNR, which may lead to a confusion here. Sometimes, $d^2$ from \eqref{SNRhom_main} is called SNR, sometimes its square root $d$ is called so. In the first case, it is the ratio of power of signal and noise, while in the second case it is the ratio of their amplitudes. In this work, we use SNR in the first sense, namely, as $d^2$.

 The integral from Eq.~\eqref{SNRhom_main} is solved numerically. The sensitivity gain in this case is a ratio of the dynamically tuned and of a reference SNR, which is presented in blue circles in Fig.~\ref{fig:snr}. Its value is of the order of 17 and slightly decreases with increasing of the source's binary mass. 

 From Eq.~\eqref{SNRhom_main}, it follows that the SNR gain is proportional to the squared average output amplitude ratio between the dynamical tuning and the stationary detection. This ratio, as it can be seen from Fig.~\ref{fig:compOut}, is approximately 4, which is consistent with the obtained value for the SNR gain.

 This number is the highest possible gain we can get from dynamical tuning. To achieve it, the displacement noise should be significantly reduced. In any realistic case with displacement noise, the gain is lower.

The increase of sensitivity  during a slow quasistationary process is higher with respect to fast nonstationary processes at the same frequencies because during the transients a part of the signal is lost. As it was shown in Sec. \ref{DynBeh}, these dynamical processes are taken into consideration in the time-domain model. To estimate the influence of these processes on the detector sensitivity, we have calculated the SNR for dynamical tuning, calculated in a quasistationary approximation, i.e.~assuming that the detector switches between the stationary states instantaneously \eqref{outQuas}. This SNR improvement is presented in Fig.~\ref{fig:snr} with green circles. It equals $\sim 19$ and, in consistence with the general speculations above, is bigger than the result of time-domain simulation and independent from the source binary mass. The difference in the shape between a quasistationary and time-domain output signals is shown in Fig.~\ref{fig:gw}(c) with red and green envelopes correspondingly. This difference is significant for some part of the signal, but due to the slow change of frequency and amplitude for most of its duration, the integral influence on the SNR is only of the order of 15\,$\%$.

\subsection{Shot-noise-limited dynamically tuned detector vs displacement-noise-limited dynamically tuned detector}
\begin{figure}
  \includegraphics [angle=0,width=\columnwidth]{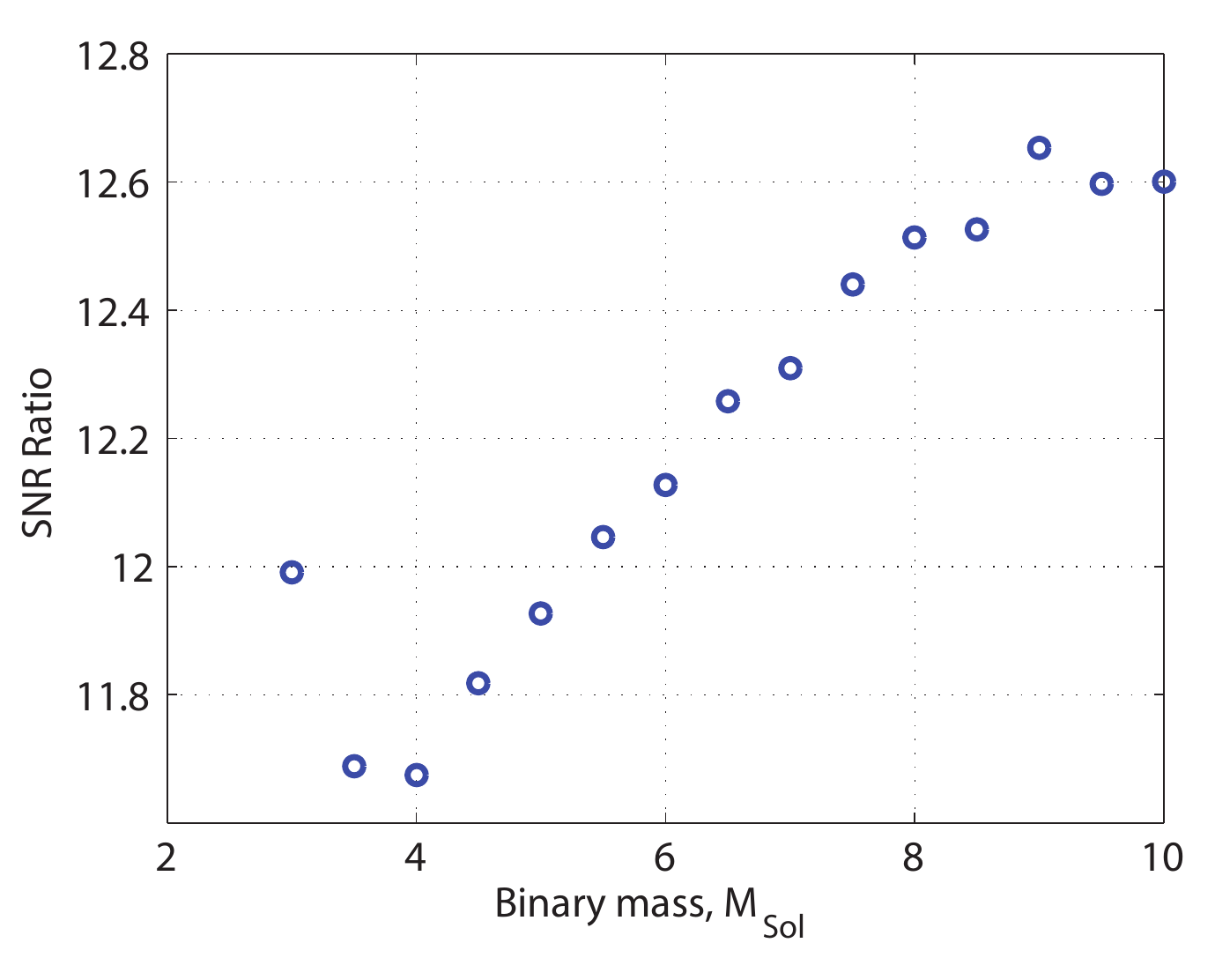}
  \caption{The ratio of SNRs for dynamical tuning in a shot-noise-limited and in a displacement-noise-limited detectors.}\label{fig:snrth}
\end{figure}
At the current operational regime, as it is shown in Fig.~\ref{fig:noises}, the displacement noise  is comparable to the shot noise in the frequency band of interest. However, since dynamical tuning dramatically decreases the influence of shot noise on the sensitivity, the influence of the displacement noise could become dominating. To check this hypothesis we compare the dynamical tuning in two special cases: (i) when the shot noise is dominating during the dynamical tuning, and we calculate the SNR consistently with Sec.~\ref{detRespSN}, using formula \eqref{SNRhom_main}, as it was performed in previous subsection, and (ii) when the displacement noise is dominating, for which we use the consideration from Sec.~\ref{detRespDN}, and find the sensitivity from Eq.~\eqref{SNR_disp}. As it was mentioned previously, the sensitivity with respect to displacement  noise is independent from the detector regime, so it will be the same for the reference detector  as  well.

The result of the comparison of the shot-noise-limited and of the displacement-noise-limited sensitivities is presented in Fig.~\ref{fig:snrth}.  The SNR for the shot-noise-dominated detector is significantly higher, approximately by factor of 12, meaning the displacement noise becomes dominating during the dynamical tuning detection.

\subsection{Displacement-noise-limited dynamically tuned detector vs reference detector with both displacement and shot noise}
The shot-noise-limited detection, considered in Sec.~\ref{subsec:snsn} is not realistic, and it is interesting to find the sensitivity improvement for the detector with the full noise budget. The radiation pressure noise is negligible, but the displacement noise is quite strong (see Fig.~\ref{fig:noises}). Dynamical tuning reduces the influence of shot noise, so the displacement noise-limited detector becomes a good approximation for this regime. We could compare the sensitivity of the displacement noise-limited dynamically tuned detector and of the broadband reference detector with the full-noise budget.

Since the sensitivity with respect to displacement noise is independent from the operational regime, this comparison is equivalent to the comparison of the broadband detector with full noise and the same detector with displacement noise only. Both SNRs may be found in the frequency domain from \eqref{SNR_disp} by substituting of the corresponding set of noise.

\begin{figure}
  \includegraphics [angle=0,width=\columnwidth]{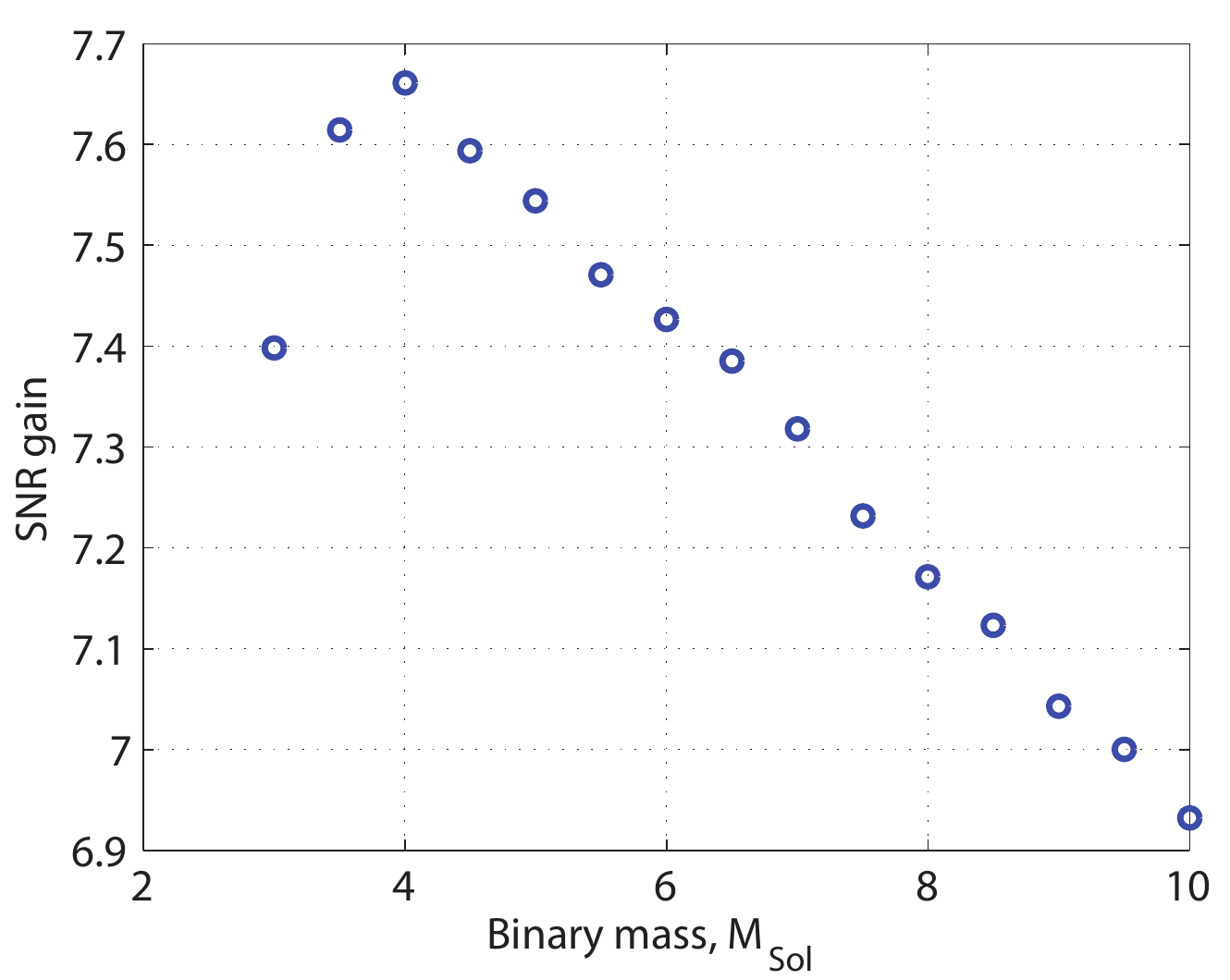}
  \caption{The SNR gain from the implementation of narrow-banded dynamical tuning into a 
broadband detector with both shot and displacement noise. Dynamical tuning is 
assumed to remove the influence of shot noise and to make the detector 
displacement noise-limited.}\label{fig:snrreal}
\end{figure}

The improvement in SNR from the implementation of dynamical tuning into the detector with displacement noise (and caused by ``removing'' the shot noise) is presented in Fig.~\ref{fig:snrreal}. It is of the order of 7, insignificantly dependent on the source binary mass. This value is the maximal possible gain from dynamical tuning performed at the detector with displacement noise. When the shot noise is not dominated by the displacement noise, the SNR gain is smaller. The reason for such strong influence of the shot noise could be, e.g.,~the error in tracking of the signal frequency by the SRM position or the use of squeezed shot noise in the referent detector.

\section{Discussion}
The set of SNRs for the dynamical tuning we presented in the previous section was obtained with very special assumptions: (i) the SRM resonantly tracks the frequency of the chirp signal \eqref{resonantCond}, and (ii) the detector is considered to be either displacement-noise or shot-noise limited. The inevitable error in the SRM position during its motion makes perfect resonant tracking of the signal impossible, preventing the signal and displacement noise from reaching their maximal amplification in comparison to shot noise. Even a small error, comparable to the bandwidth of the dynamically tuned detector, i.e.,~8\,Hz, makes the influence of shot and displacement noise of the same order.

The calculation of the SNR for both noise terms, using \eqref{SNRtime}, requires numerically solving the integral equation \eqref{qtinteq} with the composite detector noise, which can be in principle calculated with arbitrary precision,
\begin{equation}\label{eqcomp}
B_{\rm tot}(t_1,t_2) = B_{\eta}^{\rm tot}(t_1,t_2) + B_{\rm th}(t_1,t_2),
\end{equation}
where the items from the sum are taken from Eqs.~\eqref{outth} and \eqref{outcorrtot}, respectively. The solution of \eqref{qtinteq} also allows us to estimate the influence from the signal tracking error, as it was done in Ref.~\cite{meers93}, giving us the realistic benefits of the dynamical tuning.

In all the real GW detectors, dc readout is used instead of homodyne detection \cite{fricke12}. The additional leak of laser light from the power-recycling cavity, caused by the dark-fringe offset, becomes an equivalent local homodyne oscillator. The leaking power on the photodiode depends on the SRC detuning and therefore becomes time dependent during the dynamical tuning detection. The filtering of the new time-dependent ``dc" part of the photocurrent requires new solutions in the signal processing. 

The considered Michelson configuration is used only in GEO\,600, while the other GW detectors, namely, Advanced LIGO, Advanced VIRGO, and the Einstein Telescope, will have Fabry--Perot cavities in the arms. The time-domain model for their layout may be obtained by the development of the time-domain model described here. However, the shot and the displacement noise of these detectors have similar proportions as depicted in Fig.~\ref{fig:noises}; therefore, the displacement-noise-limited configurations will give a good approximation for the maximal sensitivity gain that is possible by the implementation of dynamical tuning.

\section{Conclusions}
In this paper, we have considered the problem of dynamical tuning---a particular method of detecting a chirp signal, when the GW detector is kept resonantly tuned to the instantaneous frequency of the signal via properly shifting the SRM in time.

We have developed a time-domain method of analysis since the detector performing dynamical tuning operates in a nonstationary regime (detuning of the SRC rapidly changes in time to match the frequency of the signal). 

We have considered the response of the detector to the shot noise injected through the dark port and lossy optical elements and differential motion of the end mirrors, in particular, GW signal and displacement noise. We have found that, although the optical fields describing vacuum fluctuations transform nontrivially inside the nonstationary detector, the output shot noise remains delta correlated for arbitrary realistic motions of the SRM. 

The fast changes of the signal frequency and amplitude as well as of the SRM position cause transient effects. However, by properly adjusting the mirror motion to the signal frequency, i.e.,~by performing resonant tracking, the transient effects, caused by these two parameters, are cancelled by each other, leaving only amplitude transients as dynamical effects. When the amplitude of the signal in this case changes slowly enough, a virtually stationary detection is established, and the output signal holds its stationary values at each instance, although the detection could occur far from quasistationary conditions.

Using the time-domain model, the output signals from dynamical tuning were calculated. They allowed us to give the following estimations for sensitivity improvement. Assuming a shot-noise-limited detector the enhancement factor in the SNR over the current broadband GEO\,600 configuration is 17. The influence of dynamical effects in the chirp signal detection is of the order of 15 \%. However, in the realistic case, we can neglect them because then the resonant tracking of a signal frequency makes the detector displacement noise limited, and the components of the GW signal and of the displacement noise are resonantly enhanced in the same manner (both being the differential motion of the end mirrors).  The current level of displacement noise, being considered as the sensitivity of a dynamically tuned detector (dynamical tuning removes shot noise), reduces the possible SNR enhancement factor down to 7.

The SNR for the displacment-noise-limited detector is the upper limit for the SNR gain from dynamical tuning considering the current level of theoretically predicted displacement noise. The possible reduction of this noise would increase the gain up to 17. These two values represent the idealistic cases when the dynamically tuned detector is either shot-noise or displacement-noise-limited. Taking into account the possible error of the signal frequency for the current level of displacement noise would make a sensitivity even less than 7, causing a contribution of both shot and displacement noise into the dynamical tuning sensitivity. 

\acknowledgements
 The project was paid by SFB-TR7 and Max-Planck Society for the Advancement of Science.

I want to thank the Max-Planck-Institute and personally Karsten Danzmann for providing me comfortable conditions for my scientific investigations

I would like to extend my appreciation to the people who helped me with the scientific part of this work: to Harald L\"{u}ck for supervising my job; to Sergey Tarabrin for many hours of fruitful discussions, explaining fine points in physics of GW detectors; to Denis Vasilyev for giving a glance over my work from the nonrelated-to-GW-society point of view; to Valentin Averchenko for the discussion of the mathematical issues; to Parameswaran Ajith and especially Frank Ohme for explaining the behavior of compact binaries; and to Andrzej Krolak for discussion of the dynamical tuning idea.

I would like to mention Farid Khalili, Niels L\"{o}rch, and Katherine Dooley for giving important remarks concerning this text and especially Holger Wittel and Sergey Tarabrin for putting much effort into enhancing my writing.

\begin{appendix}
\section{The impulse response of a Fabry--Perot cavity}\label{app:FP}
\begin{figure}
  \includegraphics [angle=0,width=\columnwidth]{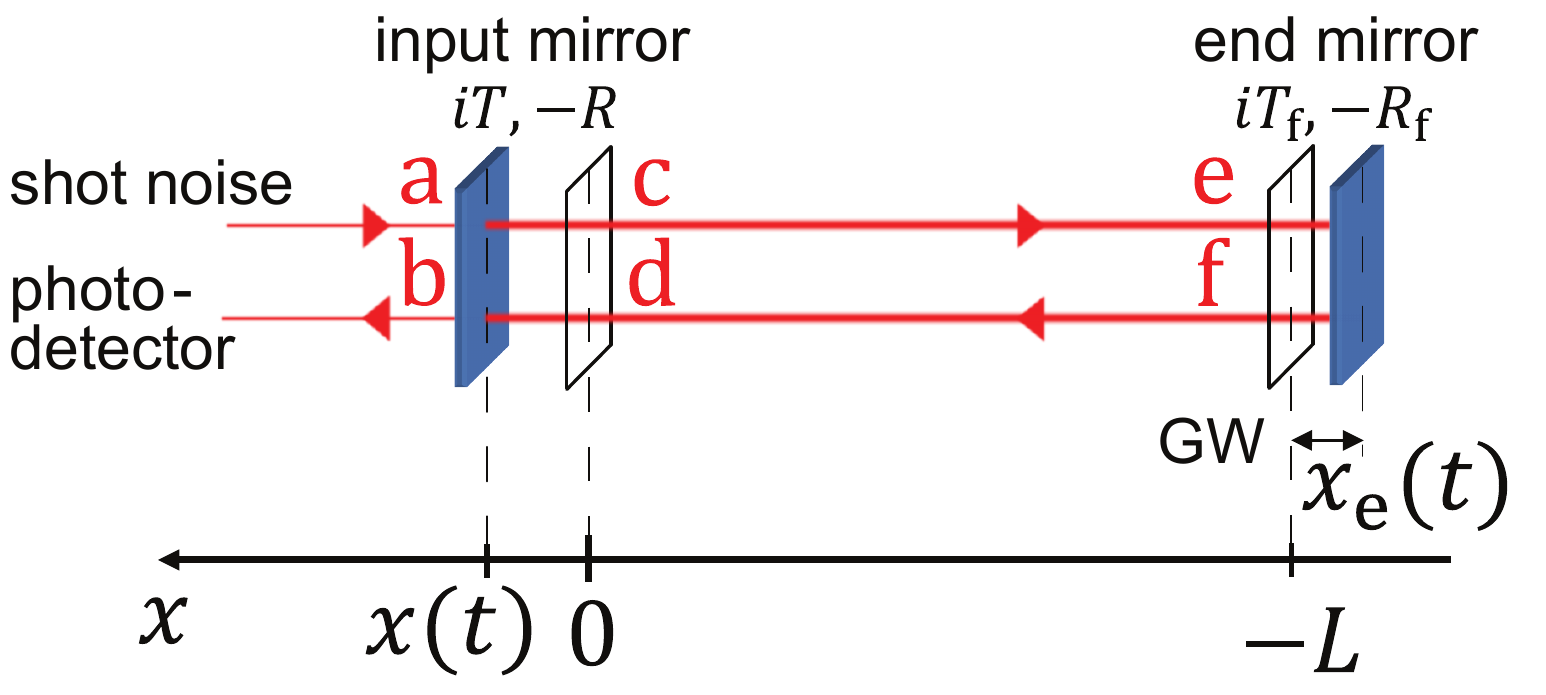}
  \caption{Scheme of the simplest Fabry--Perot cavity. $T$ and $T_{\rm f}$ are the amplitude transmittances of the input and of the end mirrors. $R$ and $R_{\rm f}$ are corresponding reflectivity coefficients. ${\rm a}$--${\rm f}$ are the electromagnetic field amplitudes in corresponding places. $L$ is the length of the cavity, resonant to laser frequency.}\label{fig:fp}
\end{figure}

A Fabry--Perot cavity (Fig.~\ref{fig:fp}) makes the simplest model for a dynamically tuned gravitational wave detector, more particularly for the SRC \cite{buonnano02} .

Plane electromagnetic waves make a good approximation for the light inside the Fabry--Perot cavity:
\begin{equation}\label{em3}
{E}_{\rm a}(t)=\sqrt{\frac{2 \pi \hbar\omega_{\rm p}}{\mathcal{A}c}}[{\rm a}(t)e^{-i\omega_{\rm p} t}+{\rm a}^\ast(t)e^{i\omega_{\rm p} t}]. 
\end{equation}
Here, $\omega_{\rm p}$ is the laser carrier frequency, $\mathcal{A}$ is the cross-section of a laser wave, and ${\rm a}(t)$ and ${\rm a}^\ast (t)$ are the complex amplitudes inside the cavity, considered in the signal range spectrum $\Omega \ll \omega_p$. The point, considered inside the cavity, is determined by denoting the field amplitude with the corresponding letter (see notations in Fig.~\ref{fig:fp}) instead of ${\rm a}(t)$.

We consider two sources of light inside the cavity: (i) the light converted from the carrier to sidebands by the differential mirror motion originates at the end mirror with the amplitude $2E k_{\rm p} x_{\rm e}(t)$ and (ii) the shot-noise injection ${\rm a}(t)$ from the input mirror, corresponding to the SRM. The complex behavior of the light inside the cavity, according to the Maxwell equations, can be described, independently from its source, described using the simple phenomena: propagation of light through the distance $x$ introducing the additional phase shift $k_{\rm p} x$, and the reflection from the mirrors with the transmission and the reflection coefficients, denoted as $\{ iT, -R\}$ (see Fig.~\ref{fig:fp}). 

We can effectively consider the field ${\rm c}(t)$ at the point of the tuned input mirror position as the result of the superposition of three different fields: (i) the input shot noise ${\rm a}(t)$ transmitted through the input mirror $i T$; (ii) the GW component injected half a round trip ago $2E k_{\rm p} x_{\rm e}(t-\tau/2)$ and reflected from the input mirror $-R$, the microscopic displacement from the resonance position of which introduces the phase shift $e^{2ik_{\rm p}x(t)}$; and (iii) the field from the same point a round trip ago ${\rm a}(t-\tau)$, propagated toward the end mirror, reflected back $-1$ (the phase shift due to GW end-mirror displacement is an effect of second order here), returned back to the input mirror and reflected from it $Re^{2ik_{\rm p}x(t)}$:
\begin{multline}\label{FP}
{\rm c}(t) = i T_{\rm S} {\rm a}(t) +2Re^{ik_{\rm p}x(t)}E k_{\rm p}x_{\rm e}(t-\tau/2) +\\
+ Re^{2ik_{\rm p}x(t)}{\rm c}(t-\tau).
\end{multline}

Now we consider the fields from the shot noise and from the GW signal separately.

\subsection{Impulse response to the GW end-mirror motion}
The fields from \eqref{FP} caused by only the gravitational wave read
\begin{multline}\label{FPGW}
{\rm c}_{\rm gw}(t) = 2Re^{ik_{\rm p}x(t)}E k_{\rm p}x_{\rm e}(t-\tau/2) +\\
+ Re^{2ik_{\rm p}x(t)}{\rm c}_{\rm gw}(t-\tau)
\end{multline}
and ``gw" stands here for ``gravitational waves."

From the solution, obtained by the recursive substitution of ${\rm c}_{\rm gw}(t)$ into the right part of the equation, the light reflected from the cavity ${\rm b}(t)$ reads 
  \begin{multline}\label{FPGW2}
{\rm b}_{\rm gw}(t) = -2iTE k_{\rm p}x_{\rm e}(t-\tau/2)- \\
-\sum\limits_{n=1}^{\infty}2iTE k_{\rm p}R_{\rm S}^n \exp[\sum\limits_{k=1}^{n}2ik_{\rm p}x(t-k\tau)]x_{\rm e}(t-b\tau-\tau/2).
\end{multline}

The photocurrent after homodyne detection with the local oscillator \eqref{LO} is
\begin{equation}\label{currentPD}
I_{\rm gw}(t) =  \sum\limits_{n=0}^{\infty}C^{\rm fp}_{n}\cos(\xi^{\rm fp}_n(t))x_{\rm e}\left(t-n\tau - \tau/2\right),
\end{equation}
where
\begin{subequations}\label{currentPDdet}
\begin{align}
C^{\rm fp}_0 &= -4\sqrt{2}\sqrt{\frac{\pi\hbar\omega_{\rm p}}{\mathcal{A}c}}T|E|k_{\rm p},\\
C^{\rm fp}_n &= C_0 R^n,\\
\xi^{\rm fp}_0(t) &= \phi_{\rm lo},\\
\xi^{\rm fp}_n(t)&= \phi_{\rm lo} + 2k_{\rm p} \sum\limits_{k=1}^nx(t-k\tau).
\end{align}
\end{subequations}

The impulse response of the detector to the GW end-mirror motion is the photocurrent, caused by a delta-impulse GW signal:
\begin{multline}\label{currentPD2}
L_{\rm s \rightarrow c}(t,t_1) =  \sum\limits_{n=0}^{\infty}C^{\rm fp}_{n}\cos(\xi^{\rm fp}_n(t))\times \\
\times \delta\left(t-t_1+n\tau + \frac{\tau}{2}\right),
\end{multline}

\subsection{Impulse response to the input shot noise}\label{repFPInp}
The quantum annihilation and creation operators of shot noise obey Eq.~\eqref{FP}, since the Maxwell equations describes the evolution of quantum fields in the same way as of the classical fields. Using the same algebraic considerations, as in the previous subsection, for the solution of \eqref{FP} with the shot-noise only influence, one gets the field on the output,
\begin{multline}\label{darkShotFieldFP} 
 {\rm b}_{\rm ip}(t)= \sum\limits_{n=1}^{\infty}T^2 \exp\left(i\varphi_n(t)\right)R^{n-1}e^{\omega_p(t_1-t)}   {\rm a}(t -n\tau)-\\
-R e^{\omega_p(t_1-t)}{\rm a} (t)\exp{i\varphi_0(t)}+h.c.,
 \end{multline}
 where 
\begin{subequations}\label{phshFP}
\begin{equation}\label{0phshFP}
\varphi_0(t) = -2k_{\rm p} x(t),
\end{equation}
\begin{equation}\label{1phshFP}
\varphi_1(t) = 0,
\end{equation}
\begin{equation}\label{nphshFP}
\varphi_n(t) = 2k_{\rm p} \sum\limits_{k=1}^{n-1} x(t-k\tau)
\end{equation}
\end{subequations}
and ``ip" stands for ``input port.''
 
By setting a deltalike impulse on the field amplitude, we get the impulse response equivalent to $L^{\rm c}(t,t_1)$, defined in  \eqref{outint}. The photocurrent of this field after the homodyne detection reads 
\begin{multline}\label{darkShotPDFP} 
 I_{\rm ip}(t)= \int\limits_{-\infty}^{t} L^{\rm c}(t,t_1)\exp\left(i\omega_{\rm p}(t-t_1)\right)z(t_1)dt_1+\\
+{\rm h.c.}
\end{multline}
Here, we introduce, equivalently to (\ref{auximp})--(\ref{phsh}):
\begin{multline}\label{LN2FP}
L_{\rm s}(t,t_1)= \sum\limits_{n=1}^{\infty}T^2 \exp\left(i\varphi_n(t)\right)R^{n-1}  \delta (t_1-t +n\tau)-\\
-R \delta (t_1-t)\exp{i\varphi_0(t)}.
\end{multline}

The autocorrelation function of the output noise may be obtained from the known input noise \eqref{SNInp}, using the impulse response (\ref{phshFP},\ref{darkShotPDFP}):
\begin{equation}\label{FPAutoCor}
B_\beta(t_1,t_2) = C_{\rm z} \delta(t_1-t_2).
\end{equation}

So, the output shot noise of the Fabry--Perot cavity with dynamically tuned SRM stays white independently from the input-mirror motion.
\subsection{Equivalent Fabry--Perot cavity}
The Fabry--Perot cavity, which is equivalent to the SRC, differs from the simplest cavity, considered in the previous subsection of the Appendix, by the nonideal end mirror with transmittance $T_{\rm f}$, equivalent to the optical losses in the cavity, and the corresponding reflectivity $R_{\rm f}$. The mirror modifies the equations for the impulse response to the signal (\ref{currentPDdet}) into
\begin{subequations}\label{currentPDdeteq}
\begin{align}
C^{\rm fp}_0 &= -4\sqrt{2}\sqrt{\frac{\pi\hbar\omega_{\rm p}}{\mathcal{A}c}}TR_{\rm f}|E|k_{\rm p},\\
C^{\rm fp}_n &= C_0 (RR_{\rm f})^n,\\
\xi^{\rm fp}_0(t) &= \phi_{\rm lo},\\
\xi^{\rm fp}_n(t)&= \phi_{\rm lo} + 2k_{\rm p} \sum\limits_{k=1}^nx(t-k\tau).
\end{align}
\end{subequations}

The equation for the auxiliary impulse response to the input mirror shot-noise injection \eqref{LN2FP} becomes:
\begin{multline}\label{LN2FPeq}
L_{\rm s}(t,t_1)= \sum\limits_{n=1}^{\infty}T^2 \exp\left(i\varphi_n(t)\right)R^{n-1}R_{\rm f}^n  \delta (t_1-t +n\tau)-\\
-R \delta (t_1-t)\exp{i\varphi_0(t)}.
\end{multline}

The additional influence of the shot noise injected into the end mirror, equivalent to the noise from the optical losses, reads
\begin{multline}\label{LN2FPem}
L_{\rm s, em} (t,t_1)= T_{\rm F} T_{\rm S} e^{i\omega_{\rm p} \frac{\tau}{2}}\times
\\ \times\sum\limits_{n=0}^{\infty}\left(R_{\rm f}R\right)^n \exp[i\varphi_{n+1}(t)]\times\\
\times  \delta\left(t_1-t+\frac{\tau}{2}+n\tau\right),
\end{multline}

\section{The impulse response of GEO\,600}\label{app:GEO}
The object of consideration in this paper is GEO\,600. Its considered layout is presented in Fig.~\ref{fig:geo1}. The actual detector has the folded arms, but for simplicity, we replace each of them by a straight arm with the same optical length. The considered parameters of the detector are presented in Tables \ref{tab:GEOconst}
and \ref{tab:GEOvar}.
In GEO\,600, we may choose four sources of the light inside the SRC: (i) the signal input of the differential motion of the end mirrors
\begin{equation}\label{xd}
x_{\rm d}(t) = \dfrac{x_{\rm e}(t)-x_{\rm n}(t)}{2};
\end{equation}
(ii) the injections of shot noise into the dark port ${\rm z}(t)$ of GEO\,600, and (iii,iv) two pieces of shot noise injected at the end mirrors ${\rm u}(t)$, ${\rm r}(t)$. 

In contrast to Appendix \ref{app:FP}, we divide the light inside interferometer into the strong part with field amplitude $A, A^\ast$, belonging to the PRC, and the weak one ${\rm a}(t), {\rm a}^\ast (t)$, belonging to the SRC:
\begin{multline}\label{em2}
E_A(t)=\sqrt{\frac{2 \pi \hbar \omega_{\rm p}}{\mathcal{A}c}} (Ae^{-i \omega_{\rm p} t}+A^\ast e^{i \omega_{\rm p} t})+\\ + \sqrt{\frac{2 \pi \hbar\omega_{\rm p}}{\mathcal{A}c}}[{\rm a}(t)e^{-i\omega_{\rm p} t}+{\rm a}^\ast(t)e^{i\omega_{\rm p} t}]. 
\end{multline}
All the other notations here are similar to those in \eqref{em3}.

\subsection{Input-output relations}

Here, the ordered input-output relations for the basic optical elements in the different arms are presented:

1. North arm (upward from the beam splitter).

a. Common mode:
\begin{subequations}\label{NP}
\begin{align}
K &= i \frac{\sqrt{2}}{2} H - \frac{\sqrt{2}}{2}E,\\
N&=Ke^{ik_{\rm p}L_{\rm n}},\\
M &= i A_{\rm n} R -R_{\rm n} N,\\
L&=Me^{ik_{\rm p}L_{\rm n}}.
\end{align}
\end{subequations}

b. Differential mode:
\begin{subequations}\label{NS}
\begin{align}
{\rm k}(t)& = i\frac{\sqrt{2}}{2} {\rm h}(t) -\frac{\sqrt{2}}{2} {\rm e}(t),\\
{\rm n}(t)&= {\rm k}(t-L_{\rm n}/c) e^{ik_{\rm p}L_{\rm n}},\\
{\rm m}(t)&=iA_{\rm n}{\rm r}(t)-R_{\rm n}{\rm n}(t) - 2iR_{\rm n}k_{\rm p}x_{\rm n}(t)N,\\
{\rm l}(t)&= {\rm m}(t-L_{\rm n}/c)e^{ik_{\rm p} L_{\rm n}}.
\end{align}
\end{subequations}

2. East arm (right-hand side from the beam splitter).

a. Common mode:
\begin{subequations}\label{EP}
\begin{align}
J &= i \frac{\sqrt{2}}{2} E - \frac{\sqrt{2}}{2}H,\\
S&=Je^{ik_{\rm p}L_{\rm e}},\\
T &= i A_{\rm e} U -R_{\rm e}S,\\
I&=Te^{ik_{\rm p}L_{\rm e}}.
\end{align}
\end{subequations}

b. Differential mode:
\begin{subequations}\label{ES}
\begin{align}
{\rm j}(t)& = i\frac{\sqrt{2}}{2} {\rm e}(t) -\frac{\sqrt{2}}{2} {\rm h}(t),\\
{\rm s}(t)&= {\rm j}(t-L_{\rm e}/c)e^{ik_{\rm p}L_{\rm e}},\\
{\rm t}(t)&=iA_{\rm e}{\rm r}(t)-R_{\rm e}{\rm s}(t) - 2iR_{\rm e}k_{\rm p}x_{\rm e}(t)S,\\
{\rm i}(t)&= {\rm t}(t-L_{\rm e}/c)e^{ik_{\rm p} L_{\rm e}}.
\end{align}
\end{subequations}

3. Signal-recycling arm (downwards from the beam splitter).

Here, we have the differential mode only:
\begin{subequations}\label{SS}
\begin{align}
{\rm g}(t)& = i\frac{\sqrt{2}}{2} {\rm l}(t) -\frac{\sqrt{2}}{2} {\rm i}(t),\\
{\rm w}(t)&= {\rm g}(t-L_{\rm s}/c)e^{ik_{\rm p }L_{\rm s}(t)},\\
{\rm o}(t)&=iT_{\rm s}{\rm z}(t)-R_{\rm s}{\rm w}(t),\\
{\rm h}(t)&= {\rm o}(t-L_{\rm s}/c)e^{ik_{\rm p }L_{\rm s}(t)},\\
{\rm y}(t) &= iT_{\rm s}{\rm w}(t)-R_{\rm s}{\rm z}(t).
\end{align}
\end{subequations}

$L_{\rm s}(t)$ here is the time-dependent distance from the SRM position to the beam splitter, setting the dynamical tuning. The change of this distance during the time of light travel between the beam splitter and the SRM is insignificant. $L_{\rm n}$ and $L_{\rm e}$ are the unperturbed lengths of the arms. The terms describing the information about the GW influence in the field reflected from the arms are obtained from the fields of the power-recycling mode by the linearization (like in, e.g., Ref.~\cite{KLMTV}). The losses in the end arms are reduced to the equivalent transmittances of the end mirrors, denoted by $A_{\rm e}$ and $A_{\rm n}$.

\subsection{Fields in the signal-recycling cavity}
Using Eqs.~(\ref{NS}), (\ref{ES}), and (\ref{SS}) and the light trajectories from Fig.~\ref{fig:geo1}, the field ${\rm h}(t)$ of the signal-recycling mode can be effectively considered as the superposition of the following rays: 

1. The shot-noise field injected through the SRM $iT_{\rm s}z(t)$. Its phase is independent from the SRM position because it is transmitted by it. For convenience we assume the phase of ${\rm h}(t)$ to be in phase with  ${\rm z}(t)$ by choosing an appropriate microscopic position of this point.

2. The field coming from the north mirror consists of the two parts: (i) the equivalent shot-noise injection due to the optical losses in the arm $iA_{\rm n}r(t)$ and (ii) the signal part carrying the information about the north end-mirror position $-2iR_{\rm n}k_{\rm  p}x_{\rm n}(t)N$. This field passes once through the north arm and the beamsplitter, followed by the reflection from the SRM with two corresponding passes through the signal-recycling arm $e^{ik_{\rm p}(L_{\rm n}+2L_{\rm s})}\times \left(i\frac{\sqrt{2}}{2}\right)\times(-R_{\rm s})$, and its time delay is  $L_{\rm n}/c+2L_{\rm s}/c$.

3. The field coming from the east mirror consists of two parts: (i) the equivalent shot-noise injection due to the optical losses in the arm $iA_{\rm e }u(t)$ and (ii) the signal part carrying the information about the east end-mirror position $-2iR_{\rm e}k_{\rm  p}x_{\rm e}(t)S$. This field passes once through the east arm and the beam splitter, followed by the reflection from the SRM with two corresponding passes through the signal-recycling arm $e^{ik_{\rm p}(L_{\rm e}+2L_{\rm s})}\times \left(-\frac{\sqrt{2}}{2}\right)\times(-R_{\rm s})$, and its time delay is  $L_{\rm e}/c+2L_{\rm s}/2$.

4. The field coming from the same point ${\rm h}(t-\tau)$ has two ways of propagation through the arms inside the SRC:

a. The part going through the north arm passes twice through the beam splitter, is once reflected from each of the north and the SRM, and twice passes through each of the north and the signal-recycling arms: $\left(i\frac{\sqrt{2}}{2}\right)\times\left(i\frac{\sqrt{2}}{2}\right)\times(-R_{\rm n})\times(-R_{\rm s})\times e^{2ik_{\rm p}(L_{\rm i}+L_{\rm s})}$. Its time delay is $2L_{\rm  s}/c + 2 L_{\rm n}/c$.

b. The part going through the east arm is reflected twice from the beam splitter, is once reflected from each of the east and the SRM and passes twice through each of the east and the output arms: $\left(-\frac{\sqrt{2}}{2}\right)\times\left(-\frac{\sqrt{2}}{2}\right)\times(-R_{\rm e})\times(-R_{\rm s})\times e^{2ik_{\rm p}(L_{\rm s}+L_{\rm e})}$. Its time delay is $2L_{\rm s}/c + 2L_{\rm e}/c$.

The light has two clearly distinguishable time evolution processes:  the microscopic change of the phase and the macroscopic time delay of the signal amplitude. The change of the phase, which is significant on the distance scales of the laser wavelength, determines the dark-port condition,
\begin{equation}\label{dPCond}
e^{i2k_{\rm p} L_{\rm n}} = -e^{i2k_{\rm p} L_{\rm e}},
\end{equation}
and the detuning of the SRM, which is taken into account in the expression $L_{\rm e}+L_{\rm s}(t) = L + x(t)$ as a microscopic displacement $x(t)$ from the length $L$  of the equivalent cavity, resonant  to the laser frequency. The delays of the signal are caused mainly be the round trips with durations $\tau = 2L/c$, while the delays, introduced by the other distance scales in this model, can be neglected.

After the construction and simplification of the expression for the ${\rm h}(t)$ considered above, we get the following expression for ${\rm{o}}(t)$:
\begin{multline}\label{GEOh112}
{\rm o}(t) \approx 2R_{\rm s}R_{\rm f}Ek_{\rm p}x_{\rm d}(t-\tau/2)e^{ik_{\rm p}x(t)}+\\
+i T_{\rm s}{\rm z}(t)e^{-ik_{\rm p}x(t)} +  i\frac{\sqrt{2}}{2}R_{\rm s} A_{\rm e} e^{ik_{\rm p}x(t) }{\rm v}(t-\tau/2 ) +\\
+i\frac{\sqrt{2}}{2}R_{\rm s} A_{\rm n} e^{ik_{\rm p}x(t)}{\rm r}(t-\tau/2)+ \\
+R_{\rm s}R_{\rm f}e^{ik_{\rm p}[x(t)+x(t-\tau)]}{\rm o}(t-\tau),
\end{multline}

where equivalent end-mirror reflectivity is
\begin{equation}\label{eqRef}
R_{\rm f} = \frac{R_{\rm e}+R_{\rm n}}{2}.
\end{equation}

The terms in its right-hand side describe the contributions during one round trip from different sources correspondingly: (i) from the signal end-mirror motion, (ii) from the shot noise injected into the dark port, (iii) from the shot noise from the losses in the east mirror, and (iv) from the shot noise from the losses in the north mirror. The fifth term of this formula describes the transformation of the field during a full round trip in the SRC.

To get the impulse response to different signal sources, we treat them separately.

\subsection{Impulse response to the differential end-mirror motion}\label{ssec:GEO_dif}
The solution for the output field amplitude ${\rm y}(t)$ obtained from the corresponding part of \eqref{GEOh112}, using \eqref{SS}, is 
\begin{multline}\label{fieldpd}
{\rm y}_{\rm dm}(t) = -2i R_{\rm f} T_{\rm s} E k_{\rm p}  x_{\rm d}\left(t-\frac{\tau}{2}\right)-\\-2i\sum\limits_{n=1}^{\infty}R_{\rm f}^{n+1}R_{\rm s}^n  T_{\rm s} E k_{\rm p}  x_{\rm d}\left(t-n\tau - \frac{\tau}{2}\right)\times \\
\times \exp \left[2ik_{\rm p} \sum\limits_{k=1}^nx(t-k\tau)\right],
\end{multline}
where ``dm" stands here for ``differential motion."

The photocurrent after the homodyne detection with the LO:
\begin{equation}\label{LO}
y_{\rm lo} = \sin(\omega_{\rm p} t + \phi_{\rm lo}),
\end{equation} 
of the field \eqref{em2} with this amplitude reads
\begin{equation}\label{currentPD3}
I_{\rm y}(t) =  \sum\limits_{n=0}^{\infty}C_{n}\cos[\xi_n(t)]x_{\rm d}\left(t-n\tau + \tau/2\right),
\end{equation}
where
\begin{subequations}\label{2pddet}
\begin{align}
C_0 &= -4\sqrt{2}\sqrt{\frac{\pi\hbar\omega_{\rm p}}{\mathcal{A}c}}R_{\rm f}T_{\rm s}|E|k_{\rm p},\\
C_n &= C_0 (R_{\rm f} R_{\rm s})^n,\\
\xi_0(t) &= \phi_{\rm h},\\
\xi_n(t)&= \phi_{\rm h} + 2k_{\rm p} \sum\limits_{k=1}^n x(t-k\tau).
\end{align}
\end{subequations}
In these expressions, $\phi_{\rm h}$ is a homodyne angle specifying the quadrature of the modulation we detect,
\begin{equation}\label{hangle}
\phi_{\rm h} = \phi_{\rm lo}+\phi_{\rm e},
\end{equation} 
where $\phi_{\rm lo}$ is a phase of the LO and $\phi_{\rm e}$ is the phase of field incident on the beam splitter $E$.

The impulse response \eqref{impulses2pd} is obtained from \eqref{currentPD3} by setting a delta impulse as the end-mirror differential motion.

For the stationary case, one can find the transfer function of the detector from the linear impulse \eqref{impulses2pd} by setting a constant detuning $2k_{\rm p} x(t) = \delta_0$ and making a Fourier transform from it:
\begin{multline}\label{TransFuncX}
R_{\rm s \rightarrow c}(\Omega) = \frac{C_0 e^{i\left(\phi_{\rm h}-\Omega\tau/2\right)}}{1-R_{\rm f}R_{\rm s}e^{i(\delta_0-\Omega\tau)}}+\\
+\frac{C_0 e^{-i\left(\phi_{\rm h}+\Omega\tau/2\right)}}{1-R_{\rm f}R_{\rm s}e^{-i(\delta_0+\Omega\tau)}}.
\end{multline}

\subsection{Inverse impulse response to the differential end-mirror motion}\label{ssec:GEO_dif_inv}
The inverse impulse response allows us to restore the motion of the end mirrors from the measured output photocurrent. It may be found from the expression for the direct impulse response \eqref{impulses2pd}. Here helps its alternative representation \eqref{detectdis1} with $M=0$, which depicts how the GW injection during the last round trip changes the field inside the cavity and, therefore, defines the difference between the latest output value and the previous one. From this expression one can derive the last injection itself and the mirror displacement causing it. The result is \eqref{invImpFunc} with the following coefficients:
\begin{subequations}\label{invImpFuncDef}
\begin{align}
\tilde{A}_0(t) &= \frac{1}{C_0},\\
\tilde{A}_1(t) &= -\frac{R}{C_0}\cos\left[2k_{\rm p}x(t-\tau+\tau/2)\right]
\end{align}
\begin{multline}
\tilde{A}_2(t) = \frac{R^2}{C_0}\sin\left[2 k_{\rm p}x(t-\tau+\tau/2)\right]\times \\
\times \sin\left[2k_{\rm p}x(t-2\tau+\tau/2)\right]
\end{multline}
\begin{multline}
\tilde{A}_n(t) = \frac{R^{n}}{C_0} \prod\limits_{l=2}^{n-1}\cos\left[2k_{\rm p}x(t-l\tau+\tau/2)\right]\times \\ 
\times \sin\left[2k_{\rm p}x(t-\tau+\tau/2)\right]\times \\
\times\sin\left[2k_{\rm p}x(t-n\tau+\tau/2)\right], n\ge 3.
\end{multline}
\end{subequations}

\subsection{Impulse response to the injection of shot noise into the dark port and into the end mirrors}\label{app:SNGEO}
The field caused by the injection of the shot noise into the dark port is obtained considering the second term and the fifth term in \eqref{fieldpd},
\begin{equation}\label{darkShotField} 
 {\rm y}_{\rm dp}(t)= \sum\limits_{n=0}^{\infty}B_n \exp\left[i\varphi_n(t)\right]e^{\omega_{\rm p}(t_1-t)} {\rm z}(t -n\tau)+h.c.,
 \end{equation}
where ``dp" stands for ``dark port" and
\begin{subequations}\label{phsh}
\begin{align}
B_0 &= -R_{\rm s},\\
B_n &= T_{\rm s}^2 R_{\rm f}^nR_{\rm s}^{n-1},\\
\varphi_0(t) &= -2k_{\rm p} x(t),\label{0phsh}\\
\varphi_1(t) &= 0,\label{1phsh}\\
\varphi_n(t) &= 2k_{\rm p} \sum\limits_{k=1}^{n-1} x(t-k\tau).\label{nphsh}
\end{align}
\end{subequations}
 The terms $\varphi_{n}(t)$ here are equivalent to those in \eqref{phshFP}.
 
The corresponding photocurrent reads
\begin{multline}\label{darkShotPD} 
 I_{\rm dp}(t)= \int\limits_{-\infty}^{t} L^{\rm c}(t,t_1)\exp\left[i\omega_{\rm p}(t-t_1)\right]z(t_1)dt_1+\\
+h.c.
 \end{multline}
 
The impulse response $L_{\rm s}(t,t_1)$ (\ref{LN2}, \ref{phsh}), defined in \eqref{auximp}, is obtained explicitly from (\ref{darkShotField}) and (\ref{darkShotPD}).

The transfer function for the field amplitude can be obtained from \eqref{LN2} by setting a constant detuning $2k_{\rm p} x(t) = \delta_0$ and Fourier transformation:
\begin{equation}\label{TransFuncDP}
R_{\rm dp}(\Omega) = e^{-i\delta_0}\left(-R_{\rm s}+\frac{T_{\rm s}^2R_{\rm f}e^{-i(\delta_0-\Omega\tau)}}{1-R_{\rm f}R_{\rm s}e^{-i(\delta_0-\Omega\tau)}}\right).
\end{equation}

In the case of the ideal end mirrors, the expression turns into
\begin{equation}\label{TransFuncDPid}
R_{\rm dp}(\Omega) = e^{-i\delta_0}\left(-R_{\rm s}+\frac{-R_{\rm s} + e^{-i(\delta_0-\Omega\tau)}}{1-R_{\rm s}e^{-i(\delta_0-\Omega\tau)}}\right).
\end{equation}

The fields caused by the injection of the shot noise through the end mirror are
\begin{subequations}
\begin{multline}\label{northShotField}
{\rm y}_{\rm nm} = A_{\rm n} T_{\rm s} \frac{\sqrt{2}}{2}\exp\left(i\omega_{\rm p} \frac{\tau}{2}\right)\sum\limits_{n=0}^{\infty}\left(R_{\rm f}R_{\rm s}\right)^n\times \\
\times\exp[i\varphi_{n+1}(t)] {\rm z}\left(t_1-t+\frac{\tau}{2}+n\tau\right)
\end{multline}
and
\begin{multline}\label{eastShotField}
{\rm y}_{\rm em} = A_{\rm e} T_{\rm s} \frac{\sqrt{2}}{2}\exp\left(i\omega_{\rm p} \frac{\tau}{2}\right)\sum\limits_{n=0}^{\infty}\left(R_{\rm f}R_{\rm s}\right)^n\times \\
\times\exp[i\varphi_{n+1}(t)] {\rm z}\left(t_1-t+\frac{\tau}{2}+n\tau\right),
\end{multline}
\end{subequations}
where ``nm" and ``em" stand for ``north mirror" and ``east mirror," correspondingly. One gets \eqref{Aimpulse2d} from this equation by substituting a delta-function input field with the following coefficients:
\begin{subequations}\label{Aimpulsecoef}
\begin{align}
B_{{\rm n}k} &= A_{\rm n} T_{\rm s} \frac{\sqrt{2}}{2}\left(R_{\rm f}R_{\rm s}\right)^k,\\
B_{{\rm e}k} &= A_{\rm e} T_{\rm s} \frac{\sqrt{2}}{2}\left(R_{\rm f}R_{\rm s}\right)^k.
\end{align}
\end{subequations}

The corresponding transfer functions, obtained from \eqref{Aimpulse2d}, for the stationary case reads
\begin{subequations}\label{Atransf}
\begin{align}
R_{\rm n}(\Omega) = \frac{\sqrt{2}}{2}\frac{A_{\rm n}T_{\rm s}}{1-R_{\rm f}R_{\rm s}e^{i(\delta-\Omega\tau)}},\\
R_{\rm e}(\Omega) = \frac{\sqrt{2}}{2}\frac{A_{\rm e}T_{\rm s}}{1-R_{\rm f}R_{\rm s}e^{i(\delta-\Omega\tau)}}.
\end{align}
\end{subequations}

\subsection{Autocorrelation function of the electromagnetic ground-state oscillations}
The shot noise of the electromagnetic field in vacuum may be measured by the joint detection of the ground-state field \eqref{em} and a local oscillator  \eqref{LO}. The result of measurement is photocurrent $I_{\rm y}$. The noise of this photocurrent is characterized by autocorrelation function:
\begin{equation}
B(t_1,t_2) \equiv \overline{I_{\rm y}(t_1)I_{\rm y}(t_2)}.
\end{equation}

The autocorrelation function for shot noise is then described by the Eq.~\eqref{SNInp} with the following coefficient:
\begin{equation}\label{SNConst}
C_{\rm z} = \frac{\pi\hbar\omega_{\rm p}}{2\mathcal{A}c}.
\end{equation}

The output shot noise in the detector is described as the response to the injection of vacuum ground-state oscillation. The autocorrelation function of such an output of the detector is then defined by the general formula \eqref{vacautocor}. The substitution of the explicit expression of the corresponding linear impulse \eqref{LN2} into it is
\begin{equation}\label{outcorrcoef}
D_n = R_{\rm f}^{|n|}T_{\rm s}^2R_{\rm s}^{|n|}\left[\dfrac{-1+R_{\rm f}^2}{1-R_{\rm f}^{2}R_{\rm s}^{2}}\right]
\end{equation}

Similarly, the output shot noise caused by the ground-state oscillations at the optical losses may be found. The result is the Eqs.~\eqref{outcorrF1}.

\subsection{Correspondence between the Fabry--Perot cavity and the GEO\,600 models}\label{equiv}

The equivalence between the time-domain models of the Fabry--Perot cavity (Fig.~\ref{fig:fp}) and of the GEO\,600 layout (Fig.~\ref{fig:geo1}) may be established based on the comparison of the impulse responses, correspondingly (\ref{currentPDdet}) and (\ref{currentPD2}) and (\ref{impulses2pd}), (\ref{2pddet}), in the following characteristic cases:

(i) the end-mirrors signal motion:
\begin{align}
x_{\rm fp}(t) &\equiv x_{\rm e}(t) = L_{\rm fp} \frac{h(t)}{2},\\
x_{\rm geo}(t) &\equiv \frac{x_{\rm e}(t)-x_{\rm n}(t)}{2} = L_{\rm geo} \frac{h(t)}{2};
\end{align}

(ii) the injection into the cavity due to the signal motion:
\begin{subequations}\label{eGW}
\begin{equation}
{\rm e}_{\rm fp}(t) \equiv { \rm e}_{\rm gw}(t) = -2iE_{\rm fp}k_{\rm p}x_{\rm fp}(t),
\end{equation}
\begin{multline}
{ \rm e}_{\rm geo}(t) \equiv \frac{{ \rm t}_{\rm gw}(t)-{ \rm m}_{\rm gw}(t)}{2} =\\
= -2iR_{\rm f} E_{\rm geo}k_{\rm p}x_{\rm geo}(t);
\end{multline}
\end{subequations}

(iii) the evolution of the fields inside the cavity during single round trips:
\begin{subequations}\label{e2GW}
\begin{align}
{\rm e}_{\rm fp}(t) &= { \rm e}_{\rm fp}(t-\tau)Re^{2ik_{\rm p} x(t-\tau/2)} ,\\
{ \rm e}_{\rm geo}(t) &= { \rm e}_{\rm geo}(t-\tau)R_SR_Fe^{2ik_{\rm p} x(t-\tau/2)};
\end{align}
\end{subequations}

(iv) the transmittance through the mirror toward the homodyne detector:
\begin{subequations}\label{yGW}
\begin{align}
{\rm y}_{\rm fp}(t) &\equiv { \rm b}_{\rm gw}(t) = iT {\rm e}_{\rm fp}(t-\tau) ,\\
{ \rm y}_{\rm geo}(t) &\equiv { \rm y}_{\rm dm}(t) = iT_{\rm S} {\rm e}_{\rm geo}(t-\tau).
\end{align}
\end{subequations}

From these relations we get the following parameters of the equivalent Fabry--Perot cavity:
\begin{subequations}\label{parFP}
\begin{align}
L_{\rm fp} &= L_{\rm geo},\\
E_{\rm fp} &= R_{\rm f}E_{\rm geo},\\
R &= R_{\rm s} R_{\rm f},\\
T & = T_{\rm s}.
\end{align}
\end{subequations} 

\section{Quasistationary approximation}\label{app:quasstat}
In the quasistationary approximation, used in this paper, we assume that all the fields at every moment of time equal their stationary values. The value for the output signal for every instance $t$ in this case could be obtained by the integrating of linear impulse \eqref{impulses2pd} with a GW signal $x_{\rm d}(t)$ with all the time-dependent parameters considered at the same instance of time.

Meers {\it et al.} have shown in Ref.~\cite{meers93}, how slow the changes of the signal parameters should be in order to keep the detector in a quasistationary regime. The boundaries are the following: (i) the amplitude and the frequency of a GW change insignificantly with respect to their absolute values during the photon lifetime inside the SRC $\tau_{\rm ph}$ [see the definition in \eqref{inSin}],
\begin{subequations}\label{domQuasyStat}
	\begin{align}
	\dot{X}(t)\tau_{\rm ph} &\ll X(t),\\
	\dot{\Omega}(t)\tau_{\rm ph} &\ll \Omega(t)
	\end{align}
	
and	(ii) the change of the detuning of the SRC during a round trip is insignificant with respect to the laser wavelength,
	\begin{equation}
	2\pi\dot{\delta}_{\rm f}(t) \tau \tau_{\rm ph}\ll 1,
	\end{equation}
\end{subequations} 
where $\delta_{\rm f}$ is a frequency detuning of the SRC from the resonance of the laser frequency and $2\pi\dot{\delta}_{\rm f}(t) \tau$ is its phase detuning.

Meers {\it et al.} state in Ref.~\cite{meers93} that for the wide variety of binary parameters the frequency $f = 500$ Hz is a reasonable upper boundary for quasistationary dynamical tuning.

The main idea of the dynamical tuning is to follow the frequency $f(t)$ of the chirp signal with the SRC detuning $\delta_{\rm f}(t)$. In a quasistationary approximation, both values can be considered as stationary at each moment of time. So, the resonant condition is
\begin{equation}\label{resConQuas}
\delta_{\rm f}(t) = f(t).
\end{equation}

If we consider the output signal under this condition, keeping only resonance terms, we get
\begin{equation}
I_{\rm y}(t) = \sum\limits_{n=0}^{\infty} \frac{1}{2}C_0 (R_{\rm f}R_{\rm s})^n x(t).
\end{equation}
This geometric series turns simply in \eqref{outQuas} with the frequency-independent coefficient:
\begin{equation}\label{Cqs}
C_{\rm qs} = \frac{1}{2}\frac{C_0}{1-R_{\rm f}R_{\rm s}}.
\end{equation}

\section{Resonant tracking}\label{app:restrack}
Now, let us assume a nonstationary case with parameters changing in time correctly. This time, we integrate the linear impulse \eqref{impulses2pd} with a chirp signal \eqref{inSin} in order to get the output signal. Under assumption of the resonance condition \eqref{resonantCond} and keeping only the resonant terms, one gets the expression
\begin{equation}
I_{\rm y}(t) = Y(t)\cos\xi(t-\tau/2),
\end{equation}
where the amplitude of the output signal is bound with the amplitude of a GW signal through:
\begin{equation}
Y(t) =  \sum\limits_{n=0}^{\infty} \frac{1}{2}C_0 (R_{\rm f}R_{\rm s})^n X(t-n\tau-\tau/2).
\end{equation}
The Fourier transform of this equation leads us to \eqref{ampTrans} with the transfer function:
\begin{equation}\label{transFunc}
R(\Omega) =  \frac{1}{2}\frac{C_0e^{i\Omega\tau/2}}{1-R_{\rm s}R_{\rm f}e^{i\Omega\tau}}.
\end{equation}
The transfer function here is an Airy function for the equivalent Fabry--Perot cavity with the frequency bandwidth determined as the round frequency $\Omega$ in \eqref{transFunc} reducing the modulus of $R(\Omega)$ by factor of $\sqrt{2}$:
\begin{equation}\label{freqBW}
\gamma = \frac{T_{\rm s}^2 + T_{\rm f}^2}{2\tau}.
\end{equation}

\section{SNR in the time-domain consideration}\label{sigDet}
The calculation of the SNR for the dynamically tuned detection of a chirp signal is based on the maximum likelihood principle, first described by Neyman and Pearson \cite{neyman33} and applied for the detection of known signals in the Gaussian noise, which is a good approximation after vetoing, e.g., in Ref.~\cite{meers932}.

Assume the two hypotheses about the measured signal $x(t)$: (i) $H_0$, assuming a pure Gaussian noise $n(t)$ with the autocorrelation function $B(t,u)$, generally speaking nonstationary,  without any signal; (ii) $H_1$, assuming the known signal $s(t)$ on the background of this noise,
\begin{equation}\label{sighyp}
x(t) = 
\begin{cases}
n(t), 0\ge t\ge T, \text{if $H_0$ is true},\\
s(t)+n(t), 0 \ge t\ge T, \text{if $H_1$ is true}.
\end{cases}
\end{equation}

For these hypotheses, the distribution of probability of measuring the discrete number of signal values $x_i = x(t_i), 0\ge i \ge N$ at the corresponding instances of time
\begin{subequations}\label{probhyp}
\begin{multline}
p_0(x_i) = \frac{1}{(2\pi)^{N/2}|S_{kl}|^{-1/2}}\times \\
\times \exp\left\{-\frac{1}{2}\sum\limits_{i,j=0}^{N}[x_i-s(t_i)]S^{-1}_{ij}[x_j-s(t_j)]\right\},
\end{multline}
\begin{equation}
p_1(x_i) = \frac{1}{(2\pi)^{N/2}|S_{kl}|^{-1/2}}\exp\left\{-\frac{1}{2}\sum\limits_{i,j=0}^{N}x_iS^{-1}_{ij}x_j\right\},
\end{equation}
\end{subequations}
where $S_{ij}\equiv  E[(x_i-s(t_i))(x_j-s(t_j))]$ is the covariation matrix that describes the noise statistics.

The likelihood ratio for this signal is
\begin{multline}\label{likelyhood}
\Lambda(x_i) \equiv \frac{p_1(x_i)}{p_0(x_i)}=\\
= \exp\left\{-\frac{1}{2}\sum\limits_{i,j=0}^{N}[x_i-s(t_i)]S^{-1}_{ij}[x_j-s(t_j)]+\right.\\
+\left.\frac{1}{2}\sum\limits_{i,j=0}^{N}x_iS^{-1}_{ij}x_j\right\}.
\end{multline}

The logarithm of likelihood for the continuous measurement may be obtained by the change of the sum over each index to the integration over the corresponding moment of time and the auxiliary substitution $q(t) = \int\limits_{0}^{T}S^{-1}(t,t_1)s(t_1)dt_1$,
\begin{equation}\label{likelyhood3}
\log\Lambda[x(t)] = \int\limits_{0}^{T}x(t)q(t)dt-\frac{1}{2}\int\limits_{0}^{T}s(t)q(t)dt,
\end{equation}
where $q(t)$ is the solution of the following integral equation:
\begin{equation}\label{qtinteq}
s(t)=\int\limits_{0}^{T}q(u)B(t,u)du.
\end{equation}


The likelihood ratio $\Lambda[x(t)]$ depends on the measured data only through an integral called a detection statistics:
\begin{equation}\label{detstat}
G = \int\limits_{0}^{T}x(t)q(t)dt.
\end{equation}

According to assumptions, every measured value $x(t)$ is Gaussian; therefore, $G$, being their linear combination, is also Gaussian, and the
parameters of its distribution are $<G> = d^2$ (for $H_1$) and $\sigma_G = <G^2-<G>^2> = d^2$, where
\begin{equation}\label{SNRtime}
d^2 = \int\limits_{0}^{T} s(t)q(t)dt
\end{equation}
is the signal-to-noise ratio.
%


In a dynamic tuning detection task, $s(t)$ is the response of the photodector current to the GW from a CBC coalescence and $B(t,u)$ is an autocorrelation function of  noise. The integral equation \eqref{qtinteq} in the case of white shot noise \eqref{outcorrtot} reads
\begin{equation}\label{qtinteqhom}
s(t)=C_{\rm z}\int\limits_{0}^{T}q(u)\delta(t-u)du = C_{\rm z} q(t),
\end{equation}
and its solution is therefore
\begin{equation}\label{qthom}
q(t)= \frac{s(t)}{C_{\rm z}}.
\end{equation}

The SNR, obtained from (\ref{SNRtime}), for the white shot noise  is
\begin{equation}\label{SNRhom}
d^2 = \frac{1}{C_{\rm z}}\int\limits_{0}^{T} s^2(t)dt.
\end{equation}

\section{The algorithm of signal simulation}\label{sigAlg}

The expression for the output \eqref{impulses2pd} consists of a very large number of summands, compared with the number of round trips during the signal detection, and therefore its  numerical calculation inevitably requires the cutoff of this sum at some rather high number. We can use the algorithm
\begin{multline}\label{detectdis1}
I_N=\sum_{n=0}^{N-1}A_n x_{{\rm d}(N-n)}\cos(\phi_N-\phi_{N-n}+\phi_{\rm h}) = \\
=\sum_{n=0}^{M}A_n x_{{\rm d}(N-n)}\cos(\phi_N-\phi_{N-n}+\phi_{\rm h}) +\\ +  R^{M+1}\left[\cos(\phi_N-\phi_{N-M-1})I_{N-M-1}-\right.\\
-\left.\sin(\phi_N-\phi_{N-M-1})I_{N-M-1(\sin)}\right]
\end{multline}
instead,  where,
\begin{subequations}
\begin{equation}
R = R_{\rm f}R_{\rm s},
\end{equation}
\begin{multline}\label{Esin}
I_{N(\sin)} = \sum_{n=0}^{N-1}A_n x_{{\rm d}(N-n)}\sin(\phi_N-\phi_{N-n}+\phi_{\rm h}) =\\
=\sum_{n=0}^{M}A_n x_{{\rm d}(N-n)}\sin(\phi_N-\phi_{N-n}+\phi_{\rm h})  + \\
+ R^{M+1}\left[\cos(\phi_N-\phi_{N-M-1})I_{N-M-1(\sin)}\right. \\
+\left.\sin(\phi_N-\phi_{N-M-1})I_{N-M-1}\right].
\end{multline}
\end{subequations}
The whole information about the infinite decaying ``tail" of the signal is used here to calculate the signal by including the phase shift, and the information about new echoes. 

The indices in \eqref{detectdis1} are chosen in the following way:
\begin{equation}
\Delta \phi_k \equiv 2k_{\rm p} x((k-1)\tau)+\phi_{\rm f}.
\end{equation}
The whole chain of phase shifts from the beginning of the measurement, cut after $j$ round trips:
\begin{equation}
\phi_j \equiv \sum_{i=1}^{j-1}\Delta \phi_i.
\end{equation}
\begin{equation}
\phi_1 = 0,
\end{equation}
\begin{subequations}\label{disc}
\begin{align}
I_N &\equiv I_{\rm y}((N-1)\tau),\\
x_{{\rm d}N}&\equiv x_{\rm d}((N-1)\tau - \tau/2).
\end{align}
\end{subequations}
The dynamical tuning resonance condition is
\begin{equation}\label{phiIndAndFInd}
\phi_n-\phi_{n-1} = 2k_{\rm p}x_{n-1} = \frac{f_{n-1}+f_{n}}{2}\tau,
\end{equation}
where
\begin{equation}
f_{N}\equiv f((N-1)\tau - \tau/2)
\end{equation}
is an instantaneous chirp frequency.
\end{appendix}

\end{document}